\documentclass[english]{jflart}
\usepackage[utf8]{inputenc}
\usepackage[T1]{fontenc}
\usepackage{xpatch}
\usepackage[frozencache]{minted}
\usepackage{ tipa }
\usepackage{float}
\usepackage{stmaryrd}
\usepackage[normalem]{ulem}
\usepackage{newunicodechar}
\usepackage{enumitem}
\usepackage{caption}
\setlist[itemize]{topsep=0.5ex,itemsep=-0.5ex,partopsep=1ex,parsep=1ex}
\usepackage{tikzit}

\tikzstyle{Normal}=[scale=0.75]
\tikzstyle{Medium}=[scale=0.55]
\tikzstyle{Small}=[scale=0.45]
\tikzstyle{Bullet}=[fill=black, draw=none, shape=circle, scale=0.35]
\tikzstyle{Grey Bullet}=[fill=gray, draw=gray, shape=circle, scale=0.35]

\tikzstyle{NormalEdge}=[-, draw=black, line width=0.75pt]
\tikzstyle{red}=[-, draw=red, line width=0.75pt]
\tikzstyle{Arrow}=[->, line width=0.75pt]
\tikzstyle{Arrow dotted}=[densely dashed,->, line width=0.75pt]
\tikzstyle{Arrow dotted grey}=[densely dashed,draw=gray, ->, line width=0.75pt]
\tikzstyle{grey}=[-, draw=gray, line width=0.75pt]

\tikzset{every picture/.style={line width=10pt}}
\newcommand{\figcomment}[1]{\textcolor{gray}{#1}}

\def\figscale{1.145}

\jfla{35}{2024}

\title{Destination-passing style programming: a Haskell implementation}

\author[1]{Thomas Bagrel}
\authorrunning{Bagrel}

\affil[1]{INRIA/LORIA, Vand\oe{}uvre-lès-Nancy, 54500, France}
\affil[1]{Tweag, Paris, 75012, France}

\makeatletter
\hypersetup{
  pdftitle={\@title},%
  pdfauthor={Thomas Bagrel <thomas.bagrel@tweag.io>},%
  pdfsubject={Computer Science - Programming Languages},%
  pdfproducer={pdfTeX},%
  pdfcreator={Thomas Bagrel <thomas.bagrel@tweag.io>},%
  pdfkeywords={Linear Types, Haskell, Destinations, DPS, GHC}
}
\makeatother

\newcommand{\mnew}[1]{\colorbox{green!50}{#1}}
\newcommand{\muline}[1]{\uline{#1}}
\newcommand{\mold}[1]{\colorbox{red!50}{#1}}
\newunicodechar{⊸}{\ensuremath{\multimap}}
\newunicodechar{→}{\ensuremath{\to}}
\newunicodechar{←}{\ensuremath{\leftarrow}}
\newunicodechar{⇒}{\ensuremath{\Rightarrow}}
\newunicodechar{⇒}{\ensuremath{\Rightarrow}}
\newunicodechar{□}{\ensuremath{{\color{red} \square}}}
\newunicodechar{¤}{}
\newunicodechar{;}{\ensuremath{\fatsemi}}
\newunicodechar{∀}{\ensuremath{\forall}}
\newunicodechar{⩴}{\hspace{-0.2ex}:\hspace{-0.65ex}:\hspace{-0.2ex}}
\makeatletter
\AtBeginEnvironment{minted}{\dontdofcolorbox}
\def\dontdofcolorbox{\renewcommand\fcolorbox[4][]{##4}}
\xpatchcmd{\inputminted}{\minted@fvset}{\minted@fvset\dontdofcolorbox}{}{}
\xpatchcmd{\mintinline}{\minted@fvset}{\minted@fvset\dontdofcolorbox}{}{} 
\makeatother

\newlength{\currentparskip}
\newenvironment{unbreakable}
{%
  \setlength{\currentparskip}{\parskip}
  \setlength{\parskip}{\currentparskip}
  \par\vspace{0.5\baselineskip}
  \noindent\begin{minipage}{\textwidth}%
    \setlength{\parskip}{\currentparskip}
}
{%
  \end{minipage}%
  \par\vspace{0.5\baselineskip}
}

\begin{document}

\maketitle

\begin{abstract}
Destination-passing style programming introduces destinations, which represent the address of a write-once memory cell. Those destinations can be passed as function parameters, and thus enable the caller of a function to keep control over memory management: the body of the called function will just be responsible of filling that memory cell. This is especially useful in functional programming languages, in which the body of a function is typically responsible for allocation of the result value.

Programming with destination in Haskell is an interesting way to improve performance of critical parts of some programs, without sacrificing memory guarantees. Indeed, thanks to a linearly-typed API I present, a write-once memory cell cannot be left uninitialized before being read, and is still disposed of by the garbage collector when it is not in use anymore, eliminating the risk of uninitialized read, memory leak, or double-free errors that can arise when memory is managed manually.

In this article, I present an implementation of destinations for Haskell, which relies on so-called compact regions. I demonstrate, in particular, a simple parser example for which the destination-based version uses 35\% less memory and time than its naive counterpart for large inputs.
\end{abstract}


\section{Introduction}

Destination-passing style (DPS) programming takes its source in the early days of imperative languages with manual memory management. In the C programming language, it's quite common for a function not to allocate memory itself for its result, but rather to receive a reference to a memory location where to write its result (often named \emph{out parameter}). In that scheme, the caller of the function has control over allocation and disposal of memory for the function result, and thus gets to choose where the latter will be written.

DPS programming is an adaptation of this idea for functional languages, based on two core concepts: having arbitrary data structures with \emph{holes} --- that is to say, memory cells that haven't been filled yet --- and \emph{destinations}, which are pointers to those holes. A destination can be passed around, as a first-class object of the language (unlike holes), and it allows remote action on its associated hole: when one \emph{fills} the destination with a value, that value is in fact written in the hole. As structures are allowed to have holes, they can be built from the root down, rather than from the leaves up. Indeed, children of a parent node no longer have to be specified when the parent node is created; they can be left empty (which leaves holes in the parent node), and added later through destinations to those holes. It is thus possible to write very natural solutions to problems for which the usual functional bottom-up building approach is ill-fitting. On top of better expressiveness, DPS programming can lead to better time or space performance for critical parts of a program, allowing for example tail-recursive map, or efficient difference lists.

That being said, DPS programming is not about giving unlimited manual control over memory or using mutations without restrictions. The existence of a destination is directly linked to the existence of an accompanying hole: we say that a destination is \emph{consumed} when it has already been used to write something in its associated hole. It must not be reused after that point, to ensure immutability and prevent a range of memory errors.

In this paper, I design a destination API whose memory safety (write-once model) is ensured through a linear type discipline. Linear type systems are based on Girard's Linear logic~\cite{girard_linear_1995}, and introduce the concept of \emph{linearity}: one can express through types that a function will \emph{consume} its argument exactly once given the function result is \emph{consumed} exactly once. Linearity helps to manage resources --- such as destinations --- that one should not forget to \emph{consume} (e.g. forgetting to fill a hole before reading a structure), but also that shouldn't be reused several times.

The Haskell programming language is equipped with support for linear types through its main compiler, \emph{GHC}, since version 9.0.1~\cite{bernardy_linear_2018}. But Haskell is also a \emph{pure} functional language, which means that side effects are usually not safe to produce outside of monadic contexts. This led me to set a slightly more refined goal: I wanted to hide impure memory effects related to destinations behind a \emph{pure} Haskell API, and make the whole safe through the linear type discipline. Although the proofs of type safety haven't been made yet, the early practical results seem to indicate that my API is safe, and its purity makes it more convenient to adopt DPS in a codebase compared to a monadic approach that would be more ``contaminating''.

\paragraph{The main contributions of this paper are}
\begin{itemize}
\item a linearly-typed API for destinations that let us build and manipulate data structures with holes while exposing a pure interface (Section~\ref{sec:api});
\item a first implementation of destinations for Haskell relying on so-called compact regions (Section~\ref{sec:implementation}), together with a performance evaluation (Section~\ref{sec:benchmark}). Implementation code is available in~\cite{custom_ghc} and~\cite{linear_dest} (specifically \mintinline{text}`src/Compact/Pure/Internal.hs` and \mintinline{text}`bench/Bench`).
\end{itemize}

\section{A short primer on linear types}\label{sec:intro-linearity}

Linear Haskell~\cite{bernardy_linear_2018} introduces the linear function arrow, \mintinline{haskellc}`a ⊸ b`, that guarantees that the argument of the function will be consumed exactly once when the result of the function is consumed exactly once. On the other hand, the regular function arrow \mintinline{haskellc}`a → b` doesn't guarantee how many times its argument will be consumed when its result is consumed once.

A value is said to be \emph{consumed once} (or \emph{consumed linearly}) when it is pattern-matched on and its sub-components are consumed once; or when it is passed as an argument to a linear function whose result is consumed once. A function is said to be \emph{consumed once} when it is applied to an argument and when the result is consumed exactly once. We say that a variable \mintinline{haskellc}`x` is \emph{used linearly} in an expression \mintinline{haskellc}`u` when consuming \mintinline{haskellc}`u` once implies consuming \mintinline{haskellc}`x` exactly once. Linearity on function arrows thus creates a chain of requirements about consumption of values, which is usually bootstrapped by using the \emph{scope function} trick, as detailed in Section~\ref{ssec:api-linearity}.

\paragraph{Unrestricted values}

Linear Haskell introduces a wrapper named \mintinline{haskellc}`Ur` which is used to indicate that a value in a linear context doesn't have to be used linearly. \mintinline{haskellc}`Ur a` is equivalent to $!a$ in linear logic, and there is an equivalence between \mintinline{haskellc}`Ur a ⊸ b` and \mintinline{haskellc}`a → b`.

The value \mintinline{haskellc}`(x, y)` is said to be consumed linearly only when both \mintinline{haskellc}`x` and \mintinline{haskellc}`y` are consumed exactly once; whereas \mintinline{haskellc}`¤Ur x` is considered to be consumed once as long as one pattern-matches on it, even if \mintinline{haskellc}`x` is not consumed exactly once after (it can be consumed several times or not at all). Conversely, both \mintinline{haskellc}`x` and \mintinline{haskellc}`y` are used linearly in \mintinline{haskellc}`(x, y)`, whereas \mintinline{haskellc}`x` is not used linearly in \mintinline{haskellc}`¤Ur x`. As a result, only values already wrapped in \mintinline{haskellc}`¤Ur` or coming from the left of a non-linear arrow can be put in another \mintinline{haskellc}`¤Ur` without breaking linearity. The only exceptions are values of types that implement the \mintinline{haskellc}`Movable` typeclass such as \mintinline{haskellc}`Int` or \mintinline{haskellc}`()`. \mintinline{haskellc}`Movable` provides \mintinline{haskellc}`move ⩴ a ⊸ Ur a` so a value can escape linearity restrictions.


\paragraph{Operators}

Some Haskell operators are often used in the rest of this article:

\mintinline{haskellc}`(<&>) ⩴ Functor ¤f ⇒ ¤f a ⊸ (a ⊸ b) ⊸ ¤f b` is the same as \mintinline{haskellc}`fmap` with the order of the arguments flipped: \mintinline{haskellc}`x <&> f = fmap f x`;

\mintinline{haskellc}`(;) ⩴ () ⊸ b ⊸ b` is used to chain a linear operation returning \mintinline{haskellc}`()` with one returning a value of type \mintinline{haskellc}`b` without breaking linearity;

\mintinline{haskellc}`Class ⇒ ...` is notation for typeclass constraints (resolved implicitly by the compiler).

\section{Motivating examples for DPS programming}\label{sec:motivating-examples}

The following subsections present three typical settings in which DPS programming brings expressiveness or performance benefits over a more traditional functional implementation.

\subsection{Efficient difference lists}\label{ssec:dlist}

Linked lists are a staple of functional programming, but they aren't efficient for concatenation, especially when the concatenation calls are nested to the left.

In an imperative context, it would be quite easy to concatenate linked lists efficiently. One just has to keep both a pointer to the root and to the last \emph{cons} cell of each list. Then, to concatenate two lists, one just has to mutate the last \emph{cons} cell of the first one to point to the root of the second list.

It isn't possible to do so in an immutable functional context though. Instead, \emph{difference lists} can be used: they are very fast to concatenate, and then to convert back into a list. They tend to emulate the idea of having a mutable (here, write-once) last \emph{cons} cell. Usually, a difference list \mintinline{haskellc}`x1 : ... : xn : □` is encoded by function \mintinline{haskellc}`\ys -> x1 : ... : xn : ys` taking a last element \mintinline{haskellc}`ys ⩴ [a]` and returning a value of type \mintinline{haskellc}`[a]` too.

With such a representation, concatenation is function composition: \mintinline{haskellc}`f1 <> f2 = f1 . f2`, and we have \mintinline{haskellc}`mempty = id`\footnote{\mintinline{haskellc}`mempty` and \mintinline{haskellc}`<>` are the usual notations for neutral element and binary operation of a monoid in Haskell.}, \mintinline{haskellc}`toList f = f []` and \mintinline{haskellc}`fromList xs = \ys → xs ++ ys`.

In DPS, instead of encoding the concept of a write-once hole with a function, we can represent the hole of type \mintinline{haskellc}`[a]` as a first-class object with a \emph{destination} of type \mintinline{haskellc}`Dest [a]`. A difference list now becomes an actual data structure in memory --- not just a pending computation --- that has two handles: one to the root of the list of type \mintinline{haskellc}`[a]`, and one to the yet-to-be-filled hole in the last cons cell, represented by the destination of type \mintinline{haskellc}`Dest [a]`.

With the function encoding, it isn't possible to read the list until a last element of type \mintinline{haskellc}`[a]` has been supplied to complete it. With the destination representation, this constraint must persist: the actual list \mintinline{haskellc}`[a]` shouldn't be readable until the accompanying destination is filled, as pattern-matching on the hole would lead to a dreaded \emph{segmentation fault}. This constraint is embodied by the \mintinline{haskellc}`Incomplete a b` type of our destination API: \mintinline{haskellc}`b` is what needs to be linearly consumed to make the \mintinline{haskellc}`a` readable. The \mintinline{haskellc}`b` side often carries the destinations of a structure. A difference list is then \mintinline{haskellc}`type DList a = Incomplete [a] (Dest [a])`: the \mintinline{haskellc}`Dest [a]` must be filled (with a \mintinline{haskellc}`[a]`) to get a readable \mintinline{haskellc}`[a]`.

The implementation of destination-backed difference lists is presented in Table~\ref{table:impl-dlist}. More details about the API primitives used by this implementation are given in Section~\ref{sec:api}. For now, it's important to note that \mintinline{haskellc}`fill` is a function taking a constructor as a type parameter (often used with \mintinline{haskellc}`@` for type parameter application, and \mintinline{haskellc}`'` to lift a constructor to a type).

\begin{table}[p]
  \small
  \begin{minted}[linenos]{haskellc}
  data [a] = {- nil constructor -} [] | {- cons constructor -} (:) a [a]
  
  type DList a = Incomplete [a] (Dest [a])
  
  alloc ⩴ DList a  -- API primitive (simplified signature w.r.t. Section 4)
  
  append ⩴ DList a ⊸ a → DList a
  append i x =
    i <&> \d → case fill @'(:) d of
      (dh, dt) → fillLeaf x dh ; dt
  
  concat ⩴ DList a ⊸ DList a ⊸ DList a
  concat i1 i2 = i1 <&> \dt1 → fillComp i2 dt1
  
  toList ⩴ DList a ⊸ [a]
  toList i = fromIncomplete_' (i <&> \dt → fill @'[] dt)
  \end{minted}
  \caption{Implementation of difference lists with destinations}
  \label{table:impl-dlist}
  \end{table}

\begin{figure}[p]\centering
  \hspace{-0.5cm}\begin{minipage}{0.3\textwidth}
    \scalebox{\figscale}{\begin{tikzpicture}
	\begin{pgfonlayer}{nodelayer}
		\node [style=Normal] (0) at (0, 0) {};
		\node [style=Normal] (1) at (4, 0) {};
		\node [style=Normal] (2) at (0, -1) {};
		\node [style=Normal] (3) at (4, -1) {};
		\node [style=Normal] (4) at (0, -3) {};
		\node [style=Normal] (5) at (4, -3) {};
		\node [style=Normal] (6) at (2, -0.5) {};
		\node [style=Normal] (7) at (2, -0.5) {\mintinline{haskellc}`¤Incomplete`};
		\node [style=Normal] (8) at (1, -1) {};
		\node [style=Normal] (9) at (1, -3) {};
		\node [style=Normal] (11) at (0.125, -2.125) {};
		\node [style=Normal] (12) at (0.875, -2.125) {};
		\node [style=Normal] (13) at (0.875, -2.875) {};
		\node [style=Normal] (14) at (0.125, -2.875) {};
		\node [style=Normal] (15) at (1.125, -2.125) {};
		\node [style=Normal] (16) at (1.125, -2.875) {};
		\node [style=Normal] (17) at (3.875, -2.125) {};
		\node [style=Normal] (18) at (3.875, -2.875) {};
		\node [style=Normal] (19) at (3.125, -2.125) {};
		\node [style=Normal] (20) at (3.125, -2.875) {};
		\node [style=Normal] (21) at (2.25, -2.5) {\mintinline{haskellc}`¤Dest`};
		\node [style=Bullet] (24) at (3.5, -2.5) {};
		\node [style=Normal] (25) at (0.5, -2.875) {};
		\node [style=Normal] (26) at (0.5, -1.5) {};
		\node [style=Small] (27) at (0.5, -1.5) {\hspace{0.3ex}\mintinline{haskellc}`⩴[Int]`};
		\node [style=Normal] (28) at (2.5, -1.5) {};
		\node [style=Small] (29) at (2.5, -1.5) {\mintinline{haskellc}`d ⩴ Dest [Int]`};
		\node [style=Medium] (30) at (2, 0.5) {\mintinline{haskellc}`alloc ⩴ Incomplete [Int] (Dest [Int])`};
	\end{pgfonlayer}
	\begin{pgfonlayer}{edgelayer}
		\draw [style=NormalEdge] (0.center) to (1.center);
		\draw [style=NormalEdge] (1.center) to (3.center);
		\draw [style=NormalEdge] (3.center) to (2.center);
		\draw [style=NormalEdge] (2.center) to (0.center);
		\draw [style=NormalEdge] (2.center) to (4.center);
		\draw [style=NormalEdge] (4.center) to (5.center);
		\draw [style=NormalEdge] (5.center) to (3.center);
		\draw [style=NormalEdge] (8.center) to (9.center);
		\draw [style=red] (11.center) to (12.center);
		\draw [style=red] (12.center) to (13.center);
		\draw [style=red] (13.center) to (14.center);
		\draw [style=red] (14.center) to (11.center);
		\draw [style=NormalEdge] (15.center) to (19.center);
		\draw [style=NormalEdge] (19.center) to (17.center);
		\draw [style=NormalEdge] (17.center) to (18.center);
		\draw [style=NormalEdge] (18.center) to (20.center);
		\draw [style=NormalEdge] (20.center) to (19.center);
		\draw [style=NormalEdge] (20.center) to (16.center);
		\draw [style=NormalEdge] (16.center) to (15.center);
		\draw [style=Arrow, bend left=75, looseness=0.75] (24) to (25.center);
	\end{pgfonlayer}
\end{tikzpicture}}
    \caption{\mintinline{haskellc}`alloc`}
    \label{fig:schema-alloc}
  \end{minipage}\hspace{0cm}%
  \begin{minipage}{0.7\textwidth}
    \scalebox{\figscale}{\input{schemas/dlist-toList.tikz}}
    \caption{Memory behavior of \mintinline{haskellc}`toList i`}
    \label{fig:schema-dlist-toList}
  \end{minipage}
\end{figure}

\begin{figure}[p]\centering
  \scalebox{\figscale}{\input{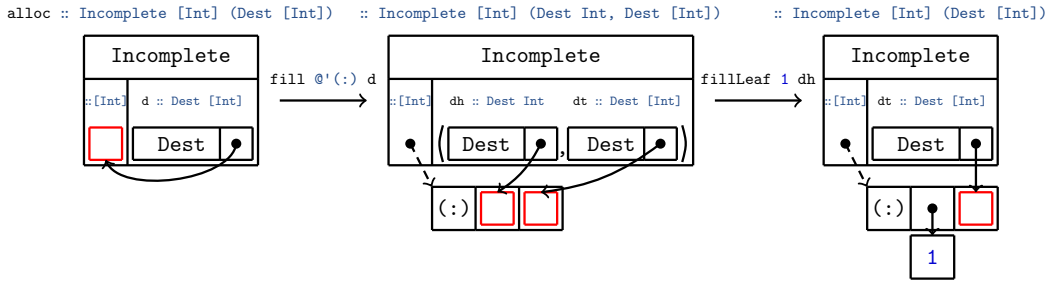}}
  \caption{Memory behavior of \mintinline{haskellc}`append alloc 1`}
  \label{fig:schema-dlist-append}
\end{figure}

\begin{figure}[p]\centering
  \scalebox{\figscale}{\input{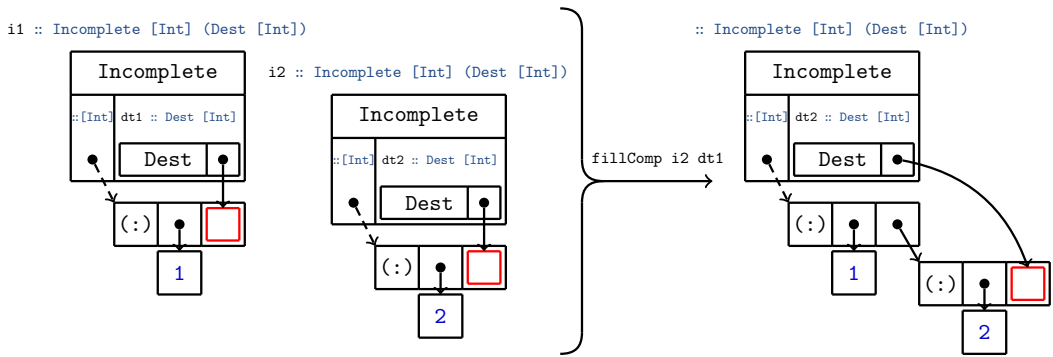}}
  \caption{Memory behavior of \mintinline{haskellc}`concat i1 i2` (based on \mintinline{haskellc}`fillComp`)}
  \label{fig:schema-dlist-concat}
\end{figure}



\begin{itemize}
  \item \mintinline{haskellc}`alloc` (Figure~\ref{fig:schema-alloc}) returns a
    \mintinline{haskellc}`DList a` which is exactly an
    \mintinline{haskellc}`Incomplete [a] (Dest [a])` structure. There is no data
    there yet and the list that will be fed in \mintinline{haskellc}`Dest [a]` is exactly the list that the
    resulting \mintinline{haskellc}`Incomplete` will hold. This is
    similar to the function encoding where \mintinline{haskellc}`\x → x` represents the empty difference list;

  \item \mintinline{haskellc}`append` (Figure~\ref{fig:schema-dlist-append}) adds an element at the tail
    position of a difference list. For this, it first uses
    \mintinline{haskellc}`fill @'(:)` to fill the hole at the end of the list represented by
    \mintinline{haskellc}`d ⩴ Dest [a]` with a hollow \emph{cons}
    cell with two new holes pointed by \mintinline{haskellc}`dh ⩴ Dest a` and \mintinline{haskellc}`dt ⩴ Dest [a]`. Then,
    \mintinline{haskellc}`fillLeaf` fills the hole represented by
    \mintinline{haskellc}`dh` with the value
    of type \mintinline{haskellc}`a`
    to append. The hole of the resulting difference list is the one pointed by \mintinline{haskellc}`dt ⩴ Dest [a]` which hasn't been filled yet.

  \item \mintinline{haskellc}`concat` (Figure~\ref{fig:schema-dlist-concat}) concatenates two difference lists,
    \mintinline{haskellc}`i1` and \mintinline{haskellc}`i2`. It uses \mintinline{haskellc}`fillComp` to fill the destination \mintinline{haskellc}`dt1`
    of the first difference list with the
    root of the second difference list \mintinline{haskellc}`i2`. The resulting \mintinline{haskellc}`Incomplete`
    object hence has the same root as the first list, holds the
    elements of both lists, and inherits the hole of the second list. Memory-wise,
    \mintinline{haskellc}`concat` just writes an address into a memory cell; no move is
    required.

  \item \mintinline{haskellc}`toList` (Figure~\ref{fig:schema-dlist-toList}) completes the incomplete structure by plugging \emph{nil} into its hole with \mintinline{haskellc}`fill @'[]` (whose individual behavior is presented in Figure~\ref{fig:schema-fillNil}) and removes the \mintinline{haskellc}`Incomplete` wrapper as the structure is now complete, using \mintinline{haskellc}`fromIncomplete_'`.
\end{itemize}

To use this API safely, it is imperative that values of type \mintinline{haskellc}`Incomplete` are used linearly. Otherwise we could first complete a difference list with \mintinline{haskellc}`l = toList i`, then add a new cons cell with a hole to \mintinline{haskellc}`i` with \mintinline{haskellc}`append i x` (actually reusing the destination inside \mintinline{haskellc}`i` for the second time). Doing that creates a hole inside \mintinline{haskellc}`l`, although it is of type \mintinline{haskellc}`[a]` so we are allowed to pattern-match on it (so we might get a segfault)! The simplified API of Table~\ref{table:impl-dlist} doesn't actually enforce the required linearity properties, I'll address that in Section~\ref{ssec:api-linearity}.

This implementation of difference list matches closely the intended memory behaviour, we can expect it to be more efficient than the functional encoding. We'll see in Section~\ref{sec:benchmark} that the prototype implementation presented in Section~\ref{sec:implementation} cannot yet demonstrate these performance improvements.

\subsection{Breadth-first tree traversal}\label{ssec:bf-tree-traversal}

Consider the problem, which Okasaki attributes to Launchbury~\cite{okasaki_bfs_2000}
\begin{quote}
  Given a tree $T$ , create a new tree of the same
  shape, but with the values at the nodes replaced
  by the numbers $1\ldots|T|$ in breadth-first order.
\end{quote}

This problem admits a straightforward implementation if we're allowed
to mutate trees. Nevertheless, a pure implementation is quite
tricky~\cite{okasaki_bfs_2000,jones_gibbons_linearbfs_93} and a very
elegant, albeit very clever, solution was proposed
recently~\cite{gibbons_phases_2023}.

With destinations as first-class objects in our toolbelt, we can
implement a solution that is both easy to come up with and efficient,
doing only a single breadth-first traversal pass on the original
tree. The main idea is to keep a queue of pairs of a tree to be
relabeled and of the destination where the relabeled result is
expected (as destinations can be stored in arbitrary containers!) and
process each of them when their turn comes. The implementation
provided in Table~\ref{table:impl-bfs-tree-traversal}, implements the
slightly more general \mintinline{haskellc}`mapAccumBFS` which applies
on each node of the tree a relabeling function that can depend on a
state.

\begin{table}[t]
\small
\begin{minted}[linenos]{haskellc}
data Tree a = ¤Nil | ¤Node a (Tree a) (Tree a)

relabelDPS ⩴ Tree a → Tree Int
relabelDPS tree = fst (mapAccumBFS (\st _ → (st + 1, st)) 1 tree)

mapAccumBFS ⩴ ∀ a b s. (s → a → (s, b)) → s → Tree a → (Tree b, s)
mapAccumBFS f s0 tree =
  fromIncomplete' (    -- simplified alloc signature w.r.t. Section 4
    alloc <&> \dtree → go s0 (singleton (¤Ur tree, dtree)))
  where
    go ⩴ s → Queue (Ur (Tree a), Dest (Tree b)) ⊸ Ur s
    go st q = case dequeue q of
      ¤Nothing → ¤Ur st
      ¤Just ((utree, dtree), q') → case utree of
        ¤Ur ¤Nil → fill @'Nil dtree ; go st q'
        ¤Ur (¤Node x tl tr) → case fill @'Node dtree of
          (dy, dtl, dtr) →
            let q'' = q' `enqueue` (¤Ur tl, dtl) `enqueue` (¤Ur tr, dtr)
                (st', y) = f st x
              in fillLeaf y dy ; go st' q''
\end{minted}
\caption{Implementation of breadth-first tree traversal with destinations}
\label{table:impl-bfs-tree-traversal}
\end{table}

Note that the signatures of \mintinline{haskellc}`mapAccumBFS` and \mintinline{haskellc}`relabelDPS` don't involve linear types. Linear types only appear in the inner loop \mintinline{haskellc}`go`, which manipulates destinations. Linearity enforces the fact that every destination ever put in the queue is eventually filled at some point, which guarantees that the output tree is complete after the function has run.

As the state-transforming function \mintinline{haskellc}`s → a → (s, b)` is non-linear, the nodes of the original tree won't be used in a linear fashion. However, we want to store these nodes together with their accompanying destinations in a queue of pairs, and destinations used to construct the queue have to be used linearly (because of \mintinline{haskellc}`<&>` signature, which initially gives the root destination). That forces the queue to be used linearly, so the pairs too, so pairs' components too. Thus, we wrap the nodes of the input tree in the \mintinline{haskellc}`Ur` wrapper, whose linear consumption allows for unrestricted use of its inner value, as detailed in Section~\ref{sec:intro-linearity}.

This example shows how destinations can be used even in a non-linear setting in order to improve the expressiveness of the language. This more natural and less convoluted implementation of breadth-first traversal also presents great performance gains compared to the fancy functional implementation from~\cite{gibbons_phases_2023}, as detailed in Section~\ref{sec:benchmark}.

\subsection{Deserializing, lifetime, and garbage collection}\label{ssec:parser-sexpr}

In client-server applications, the following pattern is very frequent: the server receives a request from a client with a serialized payload, the server then deserializes the payload, runs some code, and respond to the request. Most often, the deserialized payload is kept alive for the entirety of the request handling. In a garbage collected language, there's a real cost to this: the garbage collector (GC) will traverse the deserialized payload again and again, although we know that all its internal pointers are live for the duration of the request.

Instead, we'd rather consider the deserialized payload as a single heap object, which doesn't need to be traversed, and is freed as a block. GHC supports this use-case with a feature named \emph{compact regions}~\cite{yang_efficient_2015}. Compact regions contain normal heap objects, but the GC never follows pointers into a compact region. The flipside is that a compact region can only be collected when all of the objects it contains are dead.

With compact regions, we would first deserialize the payload normally, in the GC heap, then copy it into a compact region and only keep a reference to the copy. That way, internal pointers of the region copy will never be followed by the GC, and that copy will be collected as a whole later on, whereas the original in the GC heap will be collected immediately.

However, we are still allocating two copies of the deserialized payload. This is wasteful, it would be much better to allocate directly in the region, but this isn't part of the original compact region API. Destinations are one way to accomplish this. In fact, as I'll explain in Section~\ref{sec:implementation}, my implementation of DPS for Haskell is backed by compact regions because they provide more freedom to do low-level memory operations without interfering with GC.

Given a payload serialized as S-expressions, let's see how using destinations and compact regions for the parser can lead to greater performance. S-expressions are parenthesized lists whose elements are separated by spaces. These elements can be of several types: int, string, symbol (a textual token with no quotes around it), or a list of other S-expressions.

Parsing an S-expression can be done naively with mutually recursive functions:
\begin{itemize}
  \item \mintinline{haskellc}`parseSExpr` scans the next character, and either dispatches to \mintinline{haskellc}`parseSList` if it encounters an opening parenthesis, or to \mintinline{haskellc}`parseSString` if it encounters an opening quote, or eventually parses the string into a number or symbol;
  \item \mintinline{haskellc}`parseSList` calls \mintinline{haskellc}`parseSExpr` to parse the next token, and then calls itself again until reaching a closing parenthesis, accumulating the parsed elements along the way.
\end{itemize}

Only the implementation of \mintinline{haskellc}`parseSList` will be presented here as it is enough for our purpose, but the full implementation of both the naive and destination-based versions of the whole parser can be found in \mintinline{text}`src/Compact/Pure/SExpr.hs` of~\cite{linear_dest} .

The implementation presented in Table~\ref{table:impl-parser-naive} is quite standard: the accumulator \mintinline{haskellc}`acc` collects the nodes that are returned by \mintinline{haskellc}`parseSExpr` in the reverse order (because it's the natural building order for a linked list without destinations). When the end of the list is reached (line 5), the accumulator is reversed, wrapped in the \mintinline{haskellc}`¤SList` constructor, and returned.

\begin{table}[t]
\small
\begin{minted}[linenos,escapeinside=°°]{haskellc}
parseSList ⩴ ByteString → Int → [SExpr] → Either Error SExpr
parseSList bs i acc = case bs !? i of
  ¤Nothing → ¤Left (¤UnexpectedEOFSList i)
  ¤Just x → if
    | x == ')' → ¤Right (¤SList i (reverse acc))
    | isSpace x → parseSList bs (i + 1) acc
    | otherwise → case parseSExpr bs i of
        ¤Left err → ¤Left err
        ¤Right child → parseSList bs (endPos child + 1) (child : acc)
\end{minted}
\caption{Implementation of the S-expression parser without destinations}
\label{table:impl-parser-naive}

\bigskip

\small
\begin{minted}[linenos,escapeinside=°°]{haskellc}
parseSListDPS ⩴ ByteString → Int → Dest [SExpr] ⊸ Either Error Int
parseSListDPS bs i d = case bs !? i of
  ¤Nothing → °\mnew{fill @'[] d}° ; ¤Left (¤UnexpectedEOFSList i)
  ¤Just x → if
    | x == ')' → °\mnew{fill @'[] d}° ; ¤Right i
    | isSpace x → parseSListDPS bs (i + 1) d
    | otherwise →
        case °\mnew{fill @'(:) d}° of
          (dh, dt) -> case parseSExprDPS bs i °\mnew{dh}° of
              ¤Left err → fill @'[] dt ; ¤Left err
              ¤Right endPos → parseSListDPS bs (endPos + 1) °\mnew{dt}°
\end{minted}
\caption{Implementation of the S-expression parser with destinations}
\label{table:impl-parser-dps}
\end{table}

We will see that destinations can bring very significative performance gains with only very little stylistic changes in the code. Accumulators of tail-recursive functions just have to be changed into destinations. Instead of writing elements into a list that will be reversed at the end as we did before, the program in the destination style will directly write the elements into their final location.

Code for \mintinline{haskellc}`parseSListDPS` is presented in Table~\ref{table:impl-parser-dps}. Let's see what changed compared to the naive implementation:

\begin{itemize}
  \item even for error cases, we are forced to consume the destination that we receive as an argument (to stay linear), hence we write some sensible default data to it (see line 3);
  \item the \mintinline{haskellc}`SExpr` value resulting from \mintinline{haskellc}`parseSExprDPS` is not collected by \mintinline{haskellc}`parseSListDPS` but instead written directly into its final location by \mintinline{haskellc}`parseSExprDPS` through the passing and filling of destination \mintinline{haskellc}`dh` (see line 9);
  \item adding an element of type \mintinline{haskellc}`SExpr` to the accumulator \mintinline{haskellc}`[SExpr]` is replaced with adding a new cons cell with \mintinline{haskellc}`fill @'(:)` into the hole represented by \mintinline{haskellc}`Dest [SExpr]`, writing an element to the \emph{head} destination, and then doing a recursive call with the \emph{tail} destination passed as an argument (which has type \mintinline{haskellc}`Dest [SExpr]` again);
  \item instead of reversing and returning the accumulator at the end of the processing, it is enough to complete the list by writing a nil element to the tail destination (with \mintinline{haskellc}`fill @'[]`, see line 5), as the list has been built in a top-down approach;
  \item DPS functions return the offset of the next character to read instead of a parsed value.
\end{itemize}

Thanks to that new implementation which is barely longer (in terms of lines of code) than the naive one, the program runs almost twice as fast, mostly because garbage-collection time goes to almost zero. The detailed benchmark is available in Section~\ref{sec:benchmark}.

\section{API Design}\label{sec:api}

Table~\ref{table:destination-api} presents my pure API for functional DPS programming. This API is sufficient to implement all the examples of Section~\ref{sec:motivating-examples}. This section explains its various parts in detail.

\begin{table}[t]
\small
\begin{minted}[linenos]{haskellc}
data Token
consume   ⩴      Token ⊸ ()
dup2      ⩴      Token ⊸ (Token, Token)
withToken ⩴ ∀ a. (Token ⊸ Ur a) ⊸ Ur a

data Incomplete a b
fmap                ⩴ ∀ a b c. (b ⊸ c) ⊸ Incomplete a b ⊸ Incomplete b c
alloc               ⩴ ∀ a.     Token ⊸ Incomplete a (Dest a)
intoIncomplete      ⩴ ∀ a.     Token ⊸ a → Incomplete a ()
fromIncomplete_     ⩴ ∀ a.     Incomplete a () ⊸ Ur a
fromIncomplete      ⩴ ∀ a b.   Incomplete a (Ur c) ⊸ Ur (a, c)

data Dest a
type family DestsOf lCtor a -- returns dests associated to fields of constructor
fill     ⩴ ∀ lCtor a. Dest a ⊸ DestsOf lCtor a
fillComp ⩴ ∀ a b.     Incomplete a b ⊸ Dest a ⊸ b
fillLeaf ⩴ ∀ a.       a → Dest a ⊸ ()
\end{minted}
\caption{Destination API for Haskell}
\label{table:destination-api}
\end{table}

\subsection{The \texttt{Incomplete} type}

The main design principle behind DPS structure building is that no structure can be read before all its destinations have been filled. That way, incomplete data structures can be freely passed around and stored, but need to be completed before any pattern-matching can be made on them.

Hence we introduce a new data type \mintinline{haskellc}`Incomplete a b` where \mintinline{haskellc}`a` stands for the type of the structure being built, and \mintinline{haskellc}`b` is the type of what needs to be linearly consumed before the structure can be read. The idea is that one can map over the \mintinline{haskellc}`b` side, which will contain destinations or containers with destinations inside, until there is no destination left but just a non-linear value that can safely escape (e.g. \mintinline{haskellc}`()`, \mintinline{haskellc}`Int`, or something wrapped in \mintinline{haskellc}`Ur`). When destinations from the \mintinline{haskellc}`b` side are consumed, the structure on the \mintinline{haskellc}`a` side is built little by little in a top-down fashion, as we showed in Figures~\ref{fig:schema-dlist-append} and \ref{fig:schema-dlist-concat}. And when no destination remains on the \mintinline{haskellc}`b` side, the value of type \mintinline{haskellc}`a` no longer has holes, thus is ready to be released/read.

It can be released in two ways: with \mintinline{haskellc}`fromIncomplete_`, the value on the \mintinline{haskellc}`b` side must be unit (\mintinline{haskellc}`()`), and just the complete \mintinline{haskellc}`a` is returned, wrapped in \mintinline{haskellc}`Ur`. With \mintinline{haskellc}`fromIncomplete`, the type on the \mintinline{haskellc}`b` side must be of the form \mintinline{haskellc}`Ur c`, and then a pair \mintinline{haskellc}`Ur (a, c)` is returned.

It is actually safe to wrap the structure that has been built in \mintinline{haskellc}`Ur` because its leaves either come from non-linear sources (as \mintinline{haskellc}`fillLeaf ⩴ a → Dest a ⊸ ()` consumes its first argument non-linearly) or are made of 0-ary constructors added with \mintinline{haskellc}`fill`, both of which can be used in an unrestricted fashion safely. Variants \mintinline{haskellc}`fromIncomplete_'` and \mintinline{haskellc}`fromIncomplete'` from the beginning of this article just drop the \mintinline{haskellc}`Ur` wrapper.

Conversely, the function \mintinline{haskellc}`intoIncomplete` takes a non-linear argument of type \mintinline{haskellc}`a` and wraps it into an \mintinline{haskellc}`Incomplete` with no destinations left to be consumed.

\subsection{Ensuring write-once model for holes with linear types}\label{ssec:api-linearity}

Types aren't linear by themselves in Linear Haskell. Instead, functions can be made to use their arguments linearly or not. So in direct style, where the consumer of a resource isn't tied to the resource creation site, there is no way to state that the resource must be used exactly once:

\begin{unbreakable}
{\small
\begin{minted}[linenos]{haskellc}
createR           ⩴ Resource -- no way to force the result to be used exactly once
consumeR          ⩴ Resource ⊸ ()
exampleShouldFail ⩴ () =      
  let x = createR in consumeR x ; consumeR x -- valid even if x is consumed twice
\end{minted}
}
\end{unbreakable}

The solution is to force the consumer of a resource to become explicit at the creation site of the resource, and to check through its signature that it is indeed a linear continuation:

\begin{unbreakable}
{\small
\begin{minted}[linenos,escapeinside=°°]{haskellc}
withR       ⩴ (Resource ⊸ a) ⊸ a
consumeR    ⩴ Resource ⊸ ()
exampleFail ⩴ () = withR (\x → consumeR °\mold{x}° ; consumeR °\mold{x}°) -- not linear
\end{minted}
}
\end{unbreakable}

The \mintinline{haskellc}`Resource` type is in positive position in the signature of \mintinline{haskellc}`withR`, so that the function should somehow know how to produce a \mintinline{haskellc}`Resource`, but this is opaque for the user. What matters is that a resource can only be accessed by providing a linear continuation to \mintinline{haskellc}`withR`.

Still, this is not enough; because \mintinline{haskellc}`\x → x` is indeed a linear continuation, one could use \mintinline{haskellc}`withR (\x → x)` to leak a \mintinline{haskellc}`Resource`, and then use it in a non-linear fashion in the outside world. Hence we must forbid the resource from appearing anywhere in the return type of the continuation. To do that, we ask the return type to be wrapped in \mintinline{haskellc}`Ur`: because the resource comes from the left of a linear arrow, and doesn't implement \mintinline{haskellc}`Movable`, it cannot be wrapped in \mintinline{haskellc}`Ur` without breaking linearity (see Section~\ref{sec:intro-linearity}). On the other hand, a \mintinline{haskellc}`Movable` value of type \mintinline{haskellc}`()` or \mintinline{haskellc}`Int` can be returned:
\begin{unbreakable}
{\small
\begin{minted}[linenos,escapeinside=°°]{haskellc}
withR'       ⩴ (Resource ⊸ Ur a) ⊸ Ur a
consumeR     ⩴ Resource ⊸ ()
exampleOk'   ⩴ Ur ()       = withR' (\x → let u ⩴ () = consumeR x' in move u)
exampleFail' ⩴ Ur Resource = withR' (\x → °\mold{¤Ur x}°) -- not linear
\end{minted}
}
\end{unbreakable}

This explicit \emph{scope function} trick will no longer be necessary when linear constraints will land in GHC (see \cite{spiwack_linearly_2022}). In the meantime, this principle has been used to ensure safety of the DPS implementation in Haskell.

\paragraph{Ensuring linear use of \texttt{Incomplete} objects}

If an \mintinline{haskellc}`Incomplete` object is used linearly, then its destinations will be written to exactly once; this is ensured by the signature of \mintinline{haskellc}`fmap` for \mintinline{haskellc}`Incomplete`s. So we need to ensure that \mintinline{haskellc}`Incomplete` objects are used linearly. For that, we introduce a new type \mintinline{haskellc}`Token`. A token can be linearly exchanged one-for-one with an \mintinline{haskellc}`Incomplete` of any type with \mintinline{haskellc}`alloc`, linearly duplicated with \mintinline{haskellc}`dup2`, or linearly deleted with \mintinline{haskellc}`consume`. However, it cannot be linearly stored in \mintinline{haskellc}`Ur` as it doesn't implement \mintinline{haskellc}`Movable`.

As in the example above, we just ensure that \mintinline{haskellc}`withToken ⩴ (Token ⊸ Ur a) ⊸ Ur a` is the only source of \mintinline{haskellc}`Token`s around. Now, to produce an \mintinline{haskellc}`Incomplete`, one must get a token first, so has to be in the scope of a continuation passed to \mintinline{haskellc}`withToken`. Putting either a \mintinline{haskellc}`Token` or \mintinline{haskellc}`Incomplete` in \mintinline{haskellc}`Ur` inside the continuation would make it non-linear. So none of them can escape the scope as is, but a structure built from an \mintinline{haskellc}`Incomplete` and finalized with \mintinline{haskellc}`fromIncomplete` would be automatically wrapped in \mintinline{haskellc}`Ur`, thus could safely escape\footnote{This is why the \mintinline{haskellc}`fromIncomplete'` and \mintinline{haskellc}`fromIncomplete_'` variants aren't that useful in the actual memory-safe API (which differs slightly from the simplified examples of Section~\ref{sec:motivating-examples}): here the built structure would be stuck in the scope function without its \mintinline{haskellc}`Ur` escape pass.}.

\subsection{Filling functions for destinations}

The last part of the API is the one in charge of actually building the structures in a top-down fashion. To fill a hole represented by \mintinline{haskellc}`Dest a`, three functions are available:

\mintinline{haskellc}`fillLeaf ⩴ ∀ a. a → Dest a ⊸ ()` uses a value of type \mintinline{haskellc}`a` to fill the hole represented by the destination. The destination is consumed linearly, but the value to fill the hole isn't (as indicated by the first non-linear arrow). Memory-wise, the address of the object \mintinline{haskellc}`a` is written into the memory cell pointed to by the destination (see Figure~\ref{fig:schema-fillLeaf}).

\mintinline{haskellc}`fillComp ⩴ ∀ a b. Incomplete a b ⊸ Dest a ⊸ b` is used to plug two \mintinline{haskellc}`Incomplete` objects together. The target \mintinline{haskellc}`Incomplete` isn't represented in the signature of the function. Instead, only the target hole that will receive the address of the child is represented by \mintinline{haskellc}`Dest a`; and \mintinline{haskellc}`Incomplete a b` in the signature refers to the child object. A call to \mintinline{haskellc}`fillComp` always takes place in the scope of \mintinline{haskellc}`fmap`/\mintinline{haskellc}`<&>` over the parent object:

\begin{unbreakable}
{\small
\begin{minted}[linenos,escapeinside=°°]{haskellc}
parent ⩴ Incomplete BigStruct (Dest SmallStruct, Dest OtherStruct)
child ⩴ Incomplete SmallStruct (Dest Int)
comp = parent <&> \(ds, extra) → fillComp child ds
       ⩴ Incomplete BigStruct (Dest Int, Dest OtherStruct)
\end{minted}
}
\end{unbreakable}

The resulting structure \mintinline{haskellc}`comp` is morally a \mintinline{haskellc}`BigStruct` like \mintinline{haskellc}`parent`, that inherited the hole from the child structure (\mintinline{haskellc}`Dest Int`) and still has its other hole (\mintinline{haskellc}`Dest OtherStruct`) to be filled. An example of memory behavior of \mintinline{haskellc}`fillComp` in action can be seen in Figure~\ref{fig:schema-dlist-concat}.

\mintinline{haskellc}`fill ⩴ ∀ lCtor a. Dest a ⊸ DestsOf lCtor a` lets us build structures using layers of hollow constructors. It takes a constructor as a type parameter (\mintinline{haskellc}`lCtor`) and allocates a hollow heap object that has the same header/tag as the specified constructor but unspecified fields. The address of the allocated hollow constructor is written in the destination that is passed to \mintinline{haskellc}`fill`. As a result, one hole is now filled, but there is one new hole in the structure for each field left unspecified in the hollow constructor that is now part of the bigger structure. So \mintinline{haskellc}`fill` returns one destination of matching type for each of the fields of the constructor. An example of the memory behavior of \mintinline{haskellc}`fill @'(:) ⩴ Dest [a] ⊸ (Dest a, Dest [a])` is given in Figure~\ref{fig:schema-fillCons} and the one of \mintinline{haskellc}`fill @'[] ⩴ Dest [a] ⊸ ()` is given in Figure~\ref{fig:schema-fillNil}.

\mintinline{haskellc}`DestsOf` is a type family (i.e. a function operating on types) whose role is to map a constructor to the type of destinations for its fields. For example, \mintinline{haskellc}`DestsOf '[] [a] = ()` and \mintinline{haskellc}`DestsOf '(:) [a] = (Dest a, Dest [a])`. More generally, there is a duality between the type of a constructor \mintinline[escapeinside=°°]{haskellc}`¤Ctor ⩴ (¤f°$_1\ldots\,$°¤f°$_n$°) → a` and of the associated destination-filling functions \mintinline[escapeinside=°°]{haskellc}`fill @'Ctor ⩴ Dest a ⊸ (Dest ¤f°$_1\ldots\,$°Dest ¤f°$_n$°)`. Destination-based data building can be seen as more general than the usual bottom-up constructor approach, as we can recover \mintinline{haskellc}`¤Ctor` from the associated function \mintinline[escapeinside=°°]{haskellc}`fill @'Ctor`, but not the reverse:

\begin{unbreakable}
{\small
\begin{minted}[linenos,escapeinside=°°]{haskellc}
¤Ctor ⩴ (¤f°$_1\ldots\,$°¤f°$_n$°) → a
¤Ctor (x°$_1\ldots\,$°x°$_n$°) = fromIncomplete_' (
  alloc <&> \(d ⩴ Dest a) → case fill @'Ctor d of
    (dx°$_1\ldots\,$°dx°$_n$°) → fillLeaf x°$_1$° dx°$_1$° ; °$\ldots\,$° ; fillLeaf x°$_n$° dx°$_n$°)
\end{minted}
}
\end{unbreakable}

\begin{figure}[t]\centering
  \scalebox{\figscale}{\input{schemas/fillCons.tikz}}
  \caption{Memory behavior of \mintinline{haskellc}`fill @'(:) ⩴ Dest [a] ⊸ (Dest a, Dest [a])`}
  \label{fig:schema-fillCons}

  \scalebox{\figscale}{\begin{tikzpicture}
	\begin{pgfonlayer}{nodelayer}
		\node [style=Normal] (112) at (1.875, -1) {};
		\node [style=Normal] (113) at (1.875, -1.75) {};
		\node [style=Normal] (114) at (4.625, -1) {};
		\node [style=Normal] (115) at (4.625, -1.75) {};
		\node [style=Normal] (116) at (3.875, -1) {};
		\node [style=Normal] (117) at (3.875, -1.75) {};
		\node [style=Normal] (118) at (3, -1.375) {\mintinline{haskellc}`¤Dest`};
		\node [style=Bullet] (119) at (4.25, -1.375) {};
		\node [style=Medium] (124) at (3.25, -0.5) {\mintinline{haskellc}`d ⩴ Dest [Int]`};
		\node [style=Normal] (126) at (0.5, -2.5) {};
		\node [style=Normal] (127) at (1.5, -2.5) {};
		\node [style=Normal] (128) at (2.5, -2.5) {};
		\node [style=Normal] (129) at (2.5, -3.5) {};
		\node [style=Normal] (130) at (3.5, -2.5) {};
		\node [style=Normal] (131) at (3.5, -3.5) {};
		\node [style=Normal] (132) at (1.5, -3.5) {};
		\node [style=Normal] (133) at (0.5, -3.5) {};
		\node [style=Normal] (134) at (1, -3) {};
		\node [style=Normal] (135) at (1, -3) {\mintinline{haskellc}`(:)`};
		\node [style=Normal] (140) at (2.625, -2.625) {};
		\node [style=Normal] (141) at (3.375, -2.625) {};
		\node [style=Normal] (142) at (3.375, -3.375) {};
		\node [style=Normal] (143) at (2.625, -3.375) {};
		\node [style=Normal] (145) at (3, -2.625) {};
		\node [style=Normal] (146) at (1.5, -3.625) {};
		\node [style=Normal] (147) at (2.5, -3.625) {};
		\node [style=Normal] (148) at (2.5, -4.625) {};
		\node [style=Normal] (149) at (1.5, -4.625) {};
		\node [style=Normal] (150) at (2, -4.125) {};
		\node [style=Normal] (151) at (2, -4.125) {\mintinline{haskellc}`1`};
		\node [style=Normal] (152) at (2, -3.625) {};
		\node [style=Bullet] (153) at (2, -3) {};
		\node [style=Normal] (159) at (0, -1) {};
		\node [style=Normal] (184) at (8, -2.5) {};
		\node [style=Normal] (185) at (9, -2.5) {};
		\node [style=Normal] (186) at (10, -2.5) {};
		\node [style=Normal] (187) at (10, -3.5) {};
		\node [style=Normal] (188) at (11, -2.5) {};
		\node [style=Normal] (189) at (11, -3.5) {};
		\node [style=Normal] (190) at (9, -3.5) {};
		\node [style=Normal] (191) at (8, -3.5) {};
		\node [style=Normal] (192) at (8.5, -3) {};
		\node [style=Normal] (193) at (8.5, -3) {\mintinline{haskellc}`(:)`};
		\node [style=Normal] (199) at (9, -3.625) {};
		\node [style=Normal] (200) at (10, -3.625) {};
		\node [style=Normal] (201) at (10, -4.625) {};
		\node [style=Normal] (202) at (9, -4.625) {};
		\node [style=Normal] (203) at (9.5, -4.125) {};
		\node [style=Normal] (204) at (9.5, -4.125) {\mintinline{haskellc}`1`};
		\node [style=Normal] (205) at (9.5, -3.625) {};
		\node [style=Bullet] (206) at (9.5, -3) {};
		\node [style=Normal] (207) at (5, -3) {};
		\node [style=Normal] (208) at (7, -3) {};
		\node [style=Medium] (209) at (6, -2.5) {\mintinline{haskellc}`fill @'[] d`};
		\node [style=Normal] (210) at (10.125, -3.625) {};
		\node [style=Normal] (211) at (11.125, -3.625) {};
		\node [style=Normal] (212) at (11.125, -4.625) {};
		\node [style=Normal] (213) at (10.125, -4.625) {};
		\node [style=Normal] (214) at (10.625, -4.125) {\mintinline{haskellc}`[]`};
		\node [style=Normal] (215) at (10.625, -3.625) {};
		\node [style=Bullet] (216) at (10.5, -3) {};
		\node [style=Normal] (217) at (11.125, -3.625) {};
		\node [style=Normal] (222) at (11.125, -4.625) {};
		\node [style=Normal] (233) at (7.5, -1) {};
		\node [style=Medium] (234) at (11, -0.5) {\figcomment{no new hole}};
	\end{pgfonlayer}
	\begin{pgfonlayer}{edgelayer}
		\draw [style=NormalEdge] (112.center) to (116.center);
		\draw [style=NormalEdge] (116.center) to (114.center);
		\draw [style=NormalEdge] (114.center) to (115.center);
		\draw [style=NormalEdge] (115.center) to (117.center);
		\draw [style=NormalEdge] (117.center) to (116.center);
		\draw [style=NormalEdge] (117.center) to (113.center);
		\draw [style=NormalEdge] (113.center) to (112.center);
		\draw [style=NormalEdge] (126.center) to (127.center);
		\draw [style=NormalEdge] (127.center) to (128.center);
		\draw [style=NormalEdge] (128.center) to (130.center);
		\draw [style=NormalEdge] (130.center) to (131.center);
		\draw [style=NormalEdge] (131.center) to (129.center);
		\draw [style=NormalEdge] (129.center) to (132.center);
		\draw [style=NormalEdge] (132.center) to (133.center);
		\draw [style=NormalEdge] (133.center) to (126.center);
		\draw [style=NormalEdge] (127.center) to (132.center);
		\draw [style=NormalEdge] (128.center) to (129.center);
		\draw [style=red] (140.center) to (141.center);
		\draw [style=red] (141.center) to (142.center);
		\draw [style=red] (142.center) to (143.center);
		\draw [style=red] (143.center) to (140.center);
		\draw [style=NormalEdge] (146.center) to (147.center);
		\draw [style=NormalEdge] (148.center) to (149.center);
		\draw [style=NormalEdge] (149.center) to (146.center);
		\draw [style=NormalEdge] (147.center) to (148.center);
		\draw [style=Arrow] (153) to (152.center);
		\draw [style=Arrow, bend left=15, looseness=1.25] (119) to (145.center);
		\draw [style=Arrow dotted] (159.center) to (126.center);
		\draw [style=NormalEdge] (184.center) to (185.center);
		\draw [style=NormalEdge] (185.center) to (186.center);
		\draw [style=NormalEdge] (186.center) to (188.center);
		\draw [style=NormalEdge] (188.center) to (189.center);
		\draw [style=NormalEdge] (189.center) to (187.center);
		\draw [style=NormalEdge] (187.center) to (190.center);
		\draw [style=NormalEdge] (190.center) to (191.center);
		\draw [style=NormalEdge] (191.center) to (184.center);
		\draw [style=NormalEdge] (185.center) to (190.center);
		\draw [style=NormalEdge] (186.center) to (187.center);
		\draw [style=NormalEdge] (199.center) to (200.center);
		\draw [style=NormalEdge] (201.center) to (202.center);
		\draw [style=NormalEdge] (202.center) to (199.center);
		\draw [style=NormalEdge] (200.center) to (201.center);
		\draw [style=Arrow] (206) to (205.center);
		\draw [style=Arrow] (207.center) to (208.center);
		\draw [style=NormalEdge] (210.center) to (211.center);
		\draw [style=NormalEdge] (212.center) to (213.center);
		\draw [style=NormalEdge] (213.center) to (210.center);
		\draw [style=NormalEdge] (211.center) to (212.center);
		\draw [style=Arrow] (216) to (215.center);
		\draw [style=NormalEdge] (217.center) to (222.center);
		\draw [style=Arrow dotted] (233.center) to (184.center);
	\end{pgfonlayer}
\end{tikzpicture}}
  \caption{Memory behavior of \mintinline{haskellc}`fill @'[] ⩴ Dest [a] ⊸ ()`}
  \label{fig:schema-fillNil}

  \scalebox{\figscale}{\begin{tikzpicture}
	\begin{pgfonlayer}{nodelayer}
		\node [style=Normal] (112) at (1.875, -1) {};
		\node [style=Normal] (113) at (1.875, -1.75) {};
		\node [style=Normal] (114) at (4.625, -1) {};
		\node [style=Normal] (115) at (4.625, -1.75) {};
		\node [style=Normal] (116) at (3.875, -1) {};
		\node [style=Normal] (117) at (3.875, -1.75) {};
		\node [style=Normal] (118) at (3, -1.375) {\mintinline{haskellc}`¤Dest`};
		\node [style=Bullet] (119) at (4.25, -1.375) {};
		\node [style=Medium] (124) at (3.25, -0.5) {\mintinline{haskellc}`d ⩴ Dest Int`};
		\node [style=Normal] (126) at (0.5, -2.5) {};
		\node [style=Normal] (127) at (1.5, -2.5) {};
		\node [style=Normal] (128) at (2.5, -2.5) {};
		\node [style=Normal] (129) at (2.5, -3.5) {};
		\node [style=Normal] (130) at (3.5, -2.5) {};
		\node [style=Normal] (131) at (3.5, -3.5) {};
		\node [style=Normal] (132) at (1.5, -3.5) {};
		\node [style=Normal] (133) at (0.5, -3.5) {};
		\node [style=Normal] (134) at (1, -3) {};
		\node [style=Normal] (135) at (1, -3) {\mintinline{haskellc}`(:)`};
		\node [style=Normal] (140) at (1.625, -2.625) {};
		\node [style=Normal] (141) at (2.375, -2.625) {};
		\node [style=Normal] (142) at (2.375, -3.375) {};
		\node [style=Normal] (143) at (1.625, -3.375) {};
		\node [style=Normal] (145) at (2, -2.625) {};
		\node [style=Normal] (146) at (2.5, -3.625) {};
		\node [style=Normal] (147) at (3.5, -3.625) {};
		\node [style=Normal] (148) at (3.5, -4.625) {};
		\node [style=Normal] (149) at (2.5, -4.625) {};
		\node [style=Normal] (150) at (2, -4.125) {};
		\node [style=Normal] (151) at (3, -4.125) {\mintinline{haskellc}`[]`};
		\node [style=Normal] (152) at (3, -3.625) {};
		\node [style=Bullet] (153) at (3, -3) {};
		\node [style=Normal] (159) at (0, -1) {};
		\node [style=Normal] (184) at (8, -2.5) {};
		\node [style=Normal] (185) at (9, -2.5) {};
		\node [style=Normal] (186) at (10, -2.5) {};
		\node [style=Normal] (187) at (10, -3.5) {};
		\node [style=Normal] (188) at (11, -2.5) {};
		\node [style=Normal] (189) at (11, -3.5) {};
		\node [style=Normal] (190) at (9, -3.5) {};
		\node [style=Normal] (191) at (8, -3.5) {};
		\node [style=Normal] (192) at (8.5, -3) {};
		\node [style=Normal] (193) at (8.5, -3) {\mintinline{haskellc}`(:)`};
		\node [style=Normal] (199) at (9, -3.625) {};
		\node [style=Normal] (200) at (10, -3.625) {};
		\node [style=Normal] (201) at (10, -4.625) {};
		\node [style=Normal] (202) at (9, -4.625) {};
		\node [style=Normal] (203) at (9.5, -4.125) {};
		\node [style=Normal] (204) at (9.5, -4.125) {\mintinline{haskellc}`3`};
		\node [style=Normal] (205) at (9.5, -3.625) {};
		\node [style=Bullet] (206) at (9.5, -3) {};
		\node [style=Normal] (207) at (5, -3) {};
		\node [style=Normal] (208) at (7, -3) {};
		\node [style=Medium] (209) at (6, -2.5) {\mintinline{haskellc}`fillLeaf 3 d`};
		\node [style=Normal] (210) at (10.125, -3.625) {};
		\node [style=Normal] (211) at (11.125, -3.625) {};
		\node [style=Normal] (212) at (11.125, -4.625) {};
		\node [style=Normal] (213) at (10.125, -4.625) {};
		\node [style=Normal] (214) at (10.625, -4.125) {\mintinline{haskellc}`[]`};
		\node [style=Normal] (215) at (10.625, -3.625) {};
		\node [style=Bullet] (216) at (10.5, -3) {};
		\node [style=Normal] (217) at (11.125, -3.625) {};
		\node [style=Normal] (222) at (11.125, -4.625) {};
		\node [style=Normal] (233) at (7.5, -1) {};
		\node [style=Medium] (234) at (11, -0.5) {\figcomment{no new hole}};
	\end{pgfonlayer}
	\begin{pgfonlayer}{edgelayer}
		\draw [style=NormalEdge] (112.center) to (116.center);
		\draw [style=NormalEdge] (116.center) to (114.center);
		\draw [style=NormalEdge] (114.center) to (115.center);
		\draw [style=NormalEdge] (115.center) to (117.center);
		\draw [style=NormalEdge] (117.center) to (116.center);
		\draw [style=NormalEdge] (117.center) to (113.center);
		\draw [style=NormalEdge] (113.center) to (112.center);
		\draw [style=NormalEdge] (126.center) to (127.center);
		\draw [style=NormalEdge] (127.center) to (128.center);
		\draw [style=NormalEdge] (128.center) to (130.center);
		\draw [style=NormalEdge] (130.center) to (131.center);
		\draw [style=NormalEdge] (131.center) to (129.center);
		\draw [style=NormalEdge] (129.center) to (132.center);
		\draw [style=NormalEdge] (132.center) to (133.center);
		\draw [style=NormalEdge] (133.center) to (126.center);
		\draw [style=NormalEdge] (127.center) to (132.center);
		\draw [style=NormalEdge] (128.center) to (129.center);
		\draw [style=red] (140.center) to (141.center);
		\draw [style=red] (141.center) to (142.center);
		\draw [style=red] (142.center) to (143.center);
		\draw [style=red] (143.center) to (140.center);
		\draw [style=NormalEdge] (146.center) to (147.center);
		\draw [style=NormalEdge] (148.center) to (149.center);
		\draw [style=NormalEdge] (149.center) to (146.center);
		\draw [style=NormalEdge] (147.center) to (148.center);
		\draw [style=Arrow] (153) to (152.center);
		\draw [style=Arrow dotted] (159.center) to (126.center);
		\draw [style=NormalEdge] (184.center) to (185.center);
		\draw [style=NormalEdge] (185.center) to (186.center);
		\draw [style=NormalEdge] (186.center) to (188.center);
		\draw [style=NormalEdge] (188.center) to (189.center);
		\draw [style=NormalEdge] (189.center) to (187.center);
		\draw [style=NormalEdge] (187.center) to (190.center);
		\draw [style=NormalEdge] (190.center) to (191.center);
		\draw [style=NormalEdge] (191.center) to (184.center);
		\draw [style=NormalEdge] (185.center) to (190.center);
		\draw [style=NormalEdge] (186.center) to (187.center);
		\draw [style=NormalEdge] (199.center) to (200.center);
		\draw [style=NormalEdge] (201.center) to (202.center);
		\draw [style=NormalEdge] (202.center) to (199.center);
		\draw [style=NormalEdge] (200.center) to (201.center);
		\draw [style=Arrow] (206) to (205.center);
		\draw [style=Arrow] (207.center) to (208.center);
		\draw [style=NormalEdge] (210.center) to (211.center);
		\draw [style=NormalEdge] (212.center) to (213.center);
		\draw [style=NormalEdge] (213.center) to (210.center);
		\draw [style=NormalEdge] (211.center) to (212.center);
		\draw [style=Arrow] (216) to (215.center);
		\draw [style=NormalEdge] (217.center) to (222.center);
		\draw [style=Arrow dotted] (233.center) to (184.center);
		\draw [style=Arrow, in=90, out=-90, looseness=0.75] (119) to (145.center);
	\end{pgfonlayer}
\end{tikzpicture}}
  \caption{Memory behavior of \mintinline{haskellc}`fillLeaf ⩴ a → Dest [a] ⊸ ()`}
  \label{fig:schema-fillLeaf}
\end{figure}

\section{Implementing destinations in Haskell}\label{sec:implementation}

Having incomplete structures in the memory inherently introduces a lot of tension with both the garbage collector and compiler. Indeed, the GC assumes that every heap object it traverses is well-formed, whereas incomplete structures are absolutely ill-formed: they contain uninitialized pointers, which the GC should absolutely not follow.

The tension with the compiler is of lesser extent. The compiler can make some optimizations because it assumes that every object is immutable, while DPS programming breaks that guarantee by mutating constructors after they have been allocated (albeit only one update can happen). Fortunately, these errors are easily detected when implementing the API, and fixed by asking GHC not to inline specific parts of the code (with pragmas).

\subsection{Compact Regions}\label{ssec:impl-compact-regions}

As I teased in Section~\ref{ssec:parser-sexpr}, \emph{compact regions} from~\cite{yang_efficient_2015} make it very convenient to implement DPS programming in 
Haskell. A compact region represents a memory area in the Haskell heap that is almost fully independent from the GC and the rest of the garbage-collected heap. For the GC, each compact region is seen as a single heap object with a single lifetime. The GC can efficiently check whether there is at least one pointer in the garbage-collected heap that points into the region, and while this is the case, the region is kept alive. When this condition is no longer matched, the whole region is discarded. The result is that the GC won't traverse any node from the region: it is treated as one opaque block (even though it is actually implemented as a chain of blocks of the same size, that doesn't change the principle). Also, compact regions are immobile in memory; the GC won't move them, so a destination can just be implemented as a raw pointer (type \mintinline{haskellc}`Addr#` in Haskell): \mintinline{haskellc}`data Dest r a = ¤Dest Addr#`

By using compact regions to implement DPS programming, we completely elude the concerns of tension between the garbage collector and incomplete structures we want to build. Instead, we get two extra restrictions. First, every structure in a region must be in a fully-evaluated form. Regions are strict, and a heap object that is copied to a region is first forced into normal form. This might not always be a win; sometimes laziness, which is the default \emph{modus operandi} of the garbage-collected heap, might be preferable.

Secondly, data in a region cannot contain pointers to the garbage-collected heap, or pointers to other regions: it must be self-contained. That forces us to slightly modify the API, to add a phantom type parameter \mintinline{haskellc}`r` which tags each object with the identifier of the region it belongs to. There are two related consequences: \mintinline{haskellc}`fillLeaf` has to copy each \emph{leaf} value from the garbage-collected heap into the region in which it will be used as a leaf; and \mintinline{haskellc}`fillComp` can only plug together two \mintinline{haskellc}`Incomplete`s that come from the same region.

A typeclass \mintinline{haskellc}`Region r` is also needed to carry around the details about a region that are required for the implementation. This typeclass has a single method \mintinline{haskellc}`reflect`, not available to the user, that returns the \mintinline{haskellc}`RegionInfo` structure associated to identifier \mintinline{haskellc}`r`.

The \mintinline{haskellc}`withRegion` function is the new addition to the modified API presented in Table~\ref{table:destination-api-regions} (the \mintinline{haskellc}`Token` type and its associated functions \mintinline{haskellc}`dup2` and \mintinline{haskellc}`consume` are unchanged). \mintinline{haskellc}`withRegion` is mostly a refinement over the \mintinline{haskellc}`withToken` function from Table~\ref{table:destination-api}. It receives a continuation in which \mintinline{haskellc}`r` must be a free type variable. It then spawns both a new compact region and a fresh type \mintinline[escapeinside=°°]{haskellc}`°\muline{r}°` (not a variable), and uses the \mintinline{text}`reflection` library to provide an instance of \mintinline[escapeinside=°°]{haskellc}`Region °\muline{r}°` on-the-fly that links \mintinline[escapeinside=°°]{haskellc}`°\muline{r}°` and the \mintinline{haskellc}`RegionInfo` for the new region, and calls the continuation at type \mintinline[escapeinside=°°]{haskellc}`°\muline{r}°`. This is fairly standard practice since~\cite{launchbury_lazy_1994}.

\begin{table}[t]
\small
\begin{minted}[linenos,escapeinside=°°]{haskellc}
type Region r ⩴ Constraint
°\mnew{withRegion ⩴ ∀ a. (∀ r. Region r ⇒ Token ⊸ Ur a) ⊸ Ur a}°

data Incomplete r a b
fmap            ⩴ ∀ r a b c. (b ⊸ c) ⊸ Incomplete r a b ⊸ Incomplete r b c
alloc           ⩴ ∀ r a.     Region r ⇒ Token ⊸ Incomplete r a (Dest r a)
intoIncomplete  ⩴ ∀ r a.     Region r ⇒ Token ⊸ a → Incomplete r a ()
fromIncomplete_ ⩴ ∀ r a.     Region r ⇒ Incomplete r a () ⊸ Ur a
fromIncomplete  ⩴ ∀ r a b.   Region r ⇒ Incomplete r a (Ur c) ⊸ Ur (a, c)

data Dest r a
type family DestsOf lCtor r a
fill     ⩴ ∀ lCtor r a. Region r ⇒ Dest r a ⊸ DestsOf lCtor r a
fillComp ⩴ ∀ r a b.     Region r ⇒ Incomplete r a b ⊸ Dest r a ⊸ b
fillLeaf ⩴ ∀ r a.       Region r ⇒ a → Dest r a ⊸ ()
\end{minted}
\caption{Destination API using compact regions}
\label{table:destination-api-regions}

\medskip

\centering
  \scalebox{\figscale}{\begin{tikzpicture}
	\begin{pgfonlayer}{nodelayer}
		\node [style=Normal] (0) at (0, 0) {};
		\node [style=Normal] (1) at (4, 0) {};
		\node [style=Normal] (2) at (0, -1) {};
		\node [style=Normal] (3) at (4, -1) {};
		\node [style=Normal] (4) at (0, -3) {};
		\node [style=Normal] (5) at (4, -3) {};
		\node [style=Normal] (6) at (2, -0.5) {};
		\node [style=Normal] (7) at (2, -0.5) {\mintinline{haskellc}`¤Incomplete`};
		\node [style=Normal] (8) at (1, -1) {};
		\node [style=Normal] (9) at (1, -3) {};
		\node [style=Normal] (11) at (3.125, -4.125) {};
		\node [style=Normal] (12) at (3.875, -4.125) {};
		\node [style=Normal] (13) at (3.875, -4.875) {};
		\node [style=Normal] (14) at (3.125, -4.875) {};
		\node [style=Normal] (15) at (1.125, -2.125) {};
		\node [style=Normal] (16) at (1.125, -2.875) {};
		\node [style=Normal] (17) at (3.875, -2.125) {};
		\node [style=Normal] (18) at (3.875, -2.875) {};
		\node [style=Normal] (19) at (3.125, -2.125) {};
		\node [style=Normal] (20) at (3.125, -2.875) {};
		\node [style=Normal] (21) at (2.25, -2.5) {\mintinline{haskellc}`¤Dest`};
		\node [style=Bullet] (24) at (3.5, -2.5) {};
		\node [style=Normal] (26) at (0.5, -1.5) {};
		\node [style=Small] (27) at (0.5, -1.5) {\mintinline{haskellc}`⩴ a`};
		\node [style=Normal] (28) at (2.5, -1.5) {};
		\node [style=Small] (29) at (2.5, -1.5) {\mintinline{haskellc}`⩴ Dest r a`};
		\node [style=Medium] (30) at (2, 0.5) {\mintinline{haskellc}`alloc @r token ⩴ Incomplete r a (Dest r a)`};
		\node [style=Normal] (31) at (1, -4) {};
		\node [style=Normal] (32) at (1, -5) {};
		\node [style=Normal] (33) at (3, -4) {};
		\node [style=Normal] (34) at (3, -5) {};
		\node [style=Normal] (35) at (4, -4) {};
		\node [style=Normal] (36) at (4, -5) {};
		\node [style=Normal] (37) at (2, -4.5) {$\mathtt{IND_{\to}}$};
		\node [style=Normal] (38) at (3.5, -4.125) {};
		\node [style=Normal] (41) at (4, -5.5) {};
		\node [style=Bullet] (42) at (0.5, -2.5) {};
		\node [style=Grey Bullet] (43) at (3.5, -4.5) {};
		\node [style=Medium] (45) at (1.5, -5.55) {\figcomment{will eventually points}};
		\node [style=Medium] (47) at (1.5, -5.95) {\figcomment{to a root of type \mintinline{haskellc}`a`}};
		\node [style=Normal] (48) at (-1, -3.5) {};
		\node [style=Normal] (49) at (6, -3.5) {};
		\node [style=Medium] (50) at (5.5, -2.75) {\textcolor{red}{GC Heap}};
		\node [style=Medium] (51) at (5.5, -4.25) {\textcolor{red}{Region \mintinline{haskellc}`r`}};
		\node [style=Normal] (52) at (6.5, 0.75) {};
		\node [style=Normal] (53) at (6.5, -6.0) {};
		\node [style=Normal] (54) at (9.25, 0) {};
		\node [style=Normal] (55) at (12.25, 0) {};
		\node [style=Normal] (56) at (9.25, -1) {};
		\node [style=Normal] (57) at (12.25, -1) {};
		\node [style=Normal] (58) at (9.25, -3) {};
		\node [style=Normal] (59) at (12.25, -3) {};
		\node [style=Normal] (60) at (11.25, -0.5) {};
		\node [style=Normal] (61) at (10.75, -0.5) {\mintinline{haskellc}`¤Incomplete`};
		\node [style=Normal] (62) at (10.25, -1) {};
		\node [style=Normal] (63) at (10.25, -3) {};
		\node [style=Normal] (76) at (9.75, -1.5) {};
		\node [style=Small] (77) at (9.75, -1.5) {\mintinline{haskellc}`⩴ a`};
		\node [style=Small] (79) at (11.25, -1.5) {\mintinline{haskellc}`⩴ ()`};
		\node [style=Medium] (80) at (11.25, 0.5) {\mintinline{haskellc}`intoIncomplete @r token x ⩴ Incomplete a ()`};
		\node [style=Normal] (81) at (10.25, -4) {};
		\node [style=Normal] (82) at (10.25, -5) {};
		\node [style=Normal] (83) at (11.25, -4) {};
		\node [style=Normal] (84) at (11.25, -5) {};
		\node [style=Normal] (85) at (12.25, -4) {};
		\node [style=Normal] (86) at (12.25, -5) {};
		\node [style=Normal] (87) at (10.75, -4.5) {\mintinline{haskellc}`¤MkA`};
		\node [style=Bullet] (90) at (9.75, -2.5) {};
		\node [style=Normal] (95) at (9, -3.5) {};
		\node [style=Normal] (96) at (14.5, -3.5) {};
		\node [style=Medium] (97) at (14, -2.75) {\textcolor{red}{GC Heap}};
		\node [style=Medium] (98) at (14, -4.25) {\textcolor{red}{Region \mintinline{haskellc}`r`}};
		\node [style=Normal] (99) at (10.875, -2.125) {};
		\node [style=Normal] (100) at (11.625, -2.125) {};
		\node [style=Normal] (101) at (11.625, -2.875) {};
		\node [style=Normal] (102) at (10.875, -2.875) {};
		\node [style=Normal] (103) at (11.25, -2.5) {\mintinline{haskellc}`¤()`};
		\node [style=Normal] (104) at (11.75, -4.5) {\mintinline{haskellc}`...`};
		\node [style=Medium] (105) at (11.25, -5.5) {\mintinline{haskellc}`xCopiedInRegion ⩴ a`};
		\node [style=Normal] (107) at (8.5, -3.5) {};
		\node [style=Normal] (108) at (8.75, -3.25) {};
		\node [style=Normal] (109) at (8.75, -3.75) {};
		\node [style=Normal] (110) at (8.75, -2.25) {};
		\node [style=Normal] (111) at (9, -2) {};
		\node [style=Normal] (112) at (8.75, -4.75) {};
		\node [style=Normal] (113) at (9, -5) {};
		\node [style=Normal] (114) at (7.5, -3.25) {};
		\node [style=Medium] (115) at (7.625, -3.3) {\figcomment{no}};
		\node [style=Medium] (116) at (7.625, -3.7) {\figcomment{indirection}};
	\end{pgfonlayer}
	\begin{pgfonlayer}{edgelayer}
		\draw [style=NormalEdge] (0.center) to (1.center);
		\draw [style=NormalEdge] (1.center) to (3.center);
		\draw [style=NormalEdge] (3.center) to (2.center);
		\draw [style=NormalEdge] (2.center) to (0.center);
		\draw [style=NormalEdge] (2.center) to (4.center);
		\draw [style=NormalEdge] (4.center) to (5.center);
		\draw [style=NormalEdge] (5.center) to (3.center);
		\draw [style=NormalEdge] (8.center) to (9.center);
		\draw [style=red] (11.center) to (12.center);
		\draw [style=red] (12.center) to (13.center);
		\draw [style=red] (13.center) to (14.center);
		\draw [style=red] (14.center) to (11.center);
		\draw [style=NormalEdge] (15.center) to (19.center);
		\draw [style=NormalEdge] (19.center) to (17.center);
		\draw [style=NormalEdge] (17.center) to (18.center);
		\draw [style=NormalEdge] (18.center) to (20.center);
		\draw [style=NormalEdge] (20.center) to (19.center);
		\draw [style=NormalEdge] (20.center) to (16.center);
		\draw [style=NormalEdge] (16.center) to (15.center);
		\draw [style=NormalEdge] (31.center) to (33.center);
		\draw [style=NormalEdge] (33.center) to (35.center);
		\draw [style=NormalEdge] (35.center) to (36.center);
		\draw [style=NormalEdge] (36.center) to (34.center);
		\draw [style=NormalEdge] (34.center) to (33.center);
		\draw [style=NormalEdge] (34.center) to (32.center);
		\draw [style=NormalEdge] (32.center) to (31.center);
		\draw [style=Arrow] (24) to (38.center);
		\draw [style=Arrow dotted grey] (43) to (41.center);
		\draw [style=Arrow] (42) to (31.center);
		\draw [style=red] (48.center) to (49.center);
		\draw [style=grey] (52.center) to (53.center);
		\draw [style=NormalEdge] (54.center) to (55.center);
		\draw [style=NormalEdge] (55.center) to (57.center);
		\draw [style=NormalEdge] (57.center) to (56.center);
		\draw [style=NormalEdge] (56.center) to (54.center);
		\draw [style=NormalEdge] (56.center) to (58.center);
		\draw [style=NormalEdge] (58.center) to (59.center);
		\draw [style=NormalEdge] (59.center) to (57.center);
		\draw [style=NormalEdge] (62.center) to (63.center);
		\draw [style=NormalEdge] (81.center) to (83.center);
		\draw [style=NormalEdge] (83.center) to (85.center);
		\draw [style=NormalEdge] (85.center) to (86.center);
		\draw [style=NormalEdge] (86.center) to (84.center);
		\draw [style=NormalEdge] (84.center) to (83.center);
		\draw [style=NormalEdge] (84.center) to (82.center);
		\draw [style=NormalEdge] (82.center) to (81.center);
		\draw [style=red] (95.center) to (96.center);
		\draw [style=NormalEdge] (99.center) to (100.center);
		\draw [style=NormalEdge] (100.center) to (101.center);
		\draw [style=NormalEdge] (101.center) to (102.center);
		\draw [style=NormalEdge] (102.center) to (99.center);
		\draw [style=grey] (110.center) to (108.center);
		\draw [style=grey, bend left=45, looseness=1.25] (108.center) to (107.center);
		\draw [style=grey, bend left=45, looseness=1.25] (107.center) to (109.center);
		\draw [style=grey] (109.center) to (112.center);
		\draw [style=grey, bend right=45, looseness=1.25] (112.center) to (113.center);
		\draw [style=grey, bend right=45, looseness=1.25] (111.center) to (110.center);
		\draw [style=Arrow dotted] (90) to (81.center);
	\end{pgfonlayer}
\end{tikzpicture}}
  
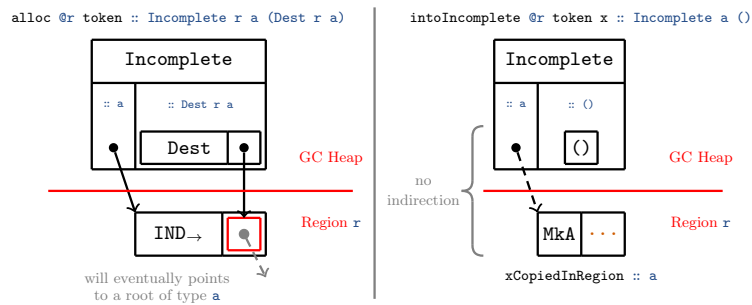
\captionof{figure}{Memory behaviour of \mintinline{haskellc}`alloc` and \mintinline{haskellc}`intoIncomplete` in the region implementation}
  \label{fig:schema-alloc-region}
\end{table}

\subsection{Representation of \texttt{Incomplete} objects}

Ideally, as we detailed in the API, we want \mintinline{haskellc}`Incomplete r a b` to contains an \mintinline{haskellc}`a` and a \mintinline{haskellc}`b`, and let the \mintinline{haskellc}`a` free when the \mintinline{haskellc}`b` is fully consumed (or linearly transformed into \mintinline{haskellc}`Ur c`). So the most straightforward implementation for \mintinline{haskellc}`Incomplete` would be a pair \mintinline{haskellc}`(a, b)`, where \mintinline{haskellc}`a` in the pair is only partially complete.

It is also natural for \mintinline{haskellc}`alloc` to return an \mintinline{haskellc}`Incomplete r a (Dest a)`: there is nothing more here than an empty memory cell (named \emph{root receiver}) of type \mintinline{haskellc}`a` which the associated destination of type \mintinline{haskellc}`Dest a` points to, as presented in Figure~\ref{fig:schema-alloc}. A bit like the identity function, whatever goes in the hole is exactly what will be retrieved in the \mintinline{haskellc}`a` side.

If \mintinline{haskellc}`Incomplete r a b` is represented by a pair \mintinline{haskellc}`(a, b)`, then the root receiver should be the first field of the pair. However, the root receiver must be in the region, otherwise the GC might follow the garbage pointer that lives inside; whereas the \mintinline{haskellc}`Incomplete` wrapper must be in the garbage-collected heap so that it can sometimes be optimized away by the compiler, and always deallocated as soon as possible.

One potential solution is to represent \mintinline{haskellc}`Incomplete r a b` by a pair \mintinline{haskellc}`(Ur a, b)` where \mintinline{haskellc}`Ur` is allocated inside the region and its field \mintinline{haskellc}`a` serves as the root receiver. With this approach, the issue of \mintinline{haskellc}`alloc` representation is solved, but every \mintinline{haskellc}`Incomplete` will now allocate a few words in the region (to host the \mintinline{haskellc}`Ur` constructor) that won't be collected by the GC for a long time even if the parent \mintinline{haskellc}`Incomplete` is collected. This makes \mintinline{haskellc}`intoIncomplete` quite inefficient memory-wise too, as the \mintinline{haskellc}`Ur` wrapper is useless for already complete structures.

The desired outcome is to only allocate a root receiver in the region for actual incomplete structures, and skip that allocation for already complete structures that are turned into an \mintinline{haskellc}`Incomplete` object, while preserving a same type for both use-cases. This is made possible by replacing the \mintinline{haskellc}`Ur` wrapper inside the \mintinline{haskellc}`Incomplete` by an indirection object (\mintinline{haskellc}`stg_IND` label) for the actually-incomplete case. \mintinline{haskellc}`Incomplete r a b` will be represented by a pair \mintinline{haskellc}`(a, b)` allocated in the garbage-collected heap, with slight variations as illustrated in Figure~\ref{fig:schema-alloc-region}:
\begin{itemize}
  \item in the pair \mintinline{haskellc}`(a, b)` returned by \mintinline{haskellc}`alloc`, the \mintinline{haskellc}`a` side points to an indirection object (a sort of constructor with one field, whose resulting type \mintinline{haskellc}`a` is the same as the field type \mintinline{haskellc}`a`), that is allocated in the region, and serves as the root receiver;
  \item in the pair \mintinline{haskellc}`(a, b)` returned by \mintinline{haskellc}`intoIncomplete`, the \mintinline{haskellc}`a` side directly points to the object of type \mintinline{haskellc}`a` that has been copied to the region.
\end{itemize}

The implementation of \mintinline{haskellc}`fromIncomplete_` is then relatively straightforward. It allocates a hollow \mintinline{haskellc}`¤Ur □` in the region, writes the address of the complete structure into it, and returns the \mintinline{haskellc}`¤Ur` (an alternative would have been to use a regular \mintinline{haskellc}`¤Ur` allocated in the GC heap).

\subsection{Deriving \texttt{fill} for all constructors with \texttt{Generics}}\label{ssec:impl-generics}

The \mintinline{haskellc}`fill @lCtor @r @a` function should plug a new hollow constructor \mintinline{haskellc}`¤Ctor □ ⩴ a` into the hole of an existing incomplete structure, and return one destination object per new hole in the structure (corresponding to the unspecified fields of the new hollow constructor). Naively, we would need one \mintinline{haskellc}`fill` function per constructor, but that cannot be realistically implemented. Instead, we have to generalize all \mintinline{haskellc}`fill` functions into a typeclass \mintinline{haskellc}`Fill lCtor a`, and derive an instance of the typeclass (i.e. implement \mintinline{haskellc}`fill`) generically for any constructor, based only on statically-known information about that constructor.

In Section~\ref{ssec:impl-ghc}, we will see how to allocate a hollow heap object for a specified constructor (which is known at compile-time). The only other information we need to implement \mintinline{haskellc}`fill` generically is the shape of the constructor, and more precisely the number and type of its fields. So we will leverage \mintinline{haskellc}`GHC.Generics` to find the required information.

\mintinline{haskellc}`GHC.Generics` is a built-in Haskell library that provides compile-time inspection of a type metadata through the \mintinline{haskellc}`Generic` typeclass: list of constructors, their fields, memory representation, etc. And that typeclass can be derived automatically for any type! Here's, for example, the \mintinline{haskellc}`Generic` representation of \mintinline{haskellc}`Maybe a`:

\begin{unbreakable}
{\small
\begin{minted}[linenos,escapeinside=°°]{haskellc}
repl> :k! Rep (Maybe a) () -- display the Generic representation of Maybe a
M1 D (MetaData "Maybe" "GHC.Maybe" "base" False) (
  M1 C (MetaCons "¤Nothing" PrefixI False) U1
  :+: M1 C (MetaCons "¤Just" PrefixI False) (M1 S ¤[...¤] (K1 R a)))
\end{minted}
}
\end{unbreakable}

We see that there are two different constructors (indicated by \mintinline{haskellc}`M1 C ...` lines): \mintinline{haskellc}`¤Nothing` has zero fields (indicated by \mintinline{haskellc}`U1`) and \mintinline{haskellc}`¤Just` has one field of type \mintinline{haskellc}`a` (indicated by \mintinline{haskellc}`K1 R a`).

With a bit of type-level programming\footnote{see \texttt{src/Compact/Pure/Internal.hs:418} in~\cite{linear_dest}}, we can extract the parts of that representation which are related to the constructor \mintinline{haskellc}`lCtor` and use them inside the instance head of \mintinline{haskellc}`Fill lCtor a` so the implementation of \mintinline{haskellc}`fill` can depend on them. That's how we can give the proper types to the destinations returned by that function for a specified constructor. The \mintinline{haskellc}`DestsOf lCtor a ⩴ Type` type family also uses the generic representation of \mintinline{haskellc}`a` to extract what it needs to know about \mintinline{haskellc}`lCtor` and its fields.

\subsection{Changes to GHC internals and RTS}\label{ssec:impl-ghc}

We will see here how to allocate a hollow heap object for a given constructor, but let's first take a detour to give more context about the internals of the compiler.

Haskell's runtime system (RTS) is written in a mix of C and C-{}-. The RTS has many roles, among which managing threads, organizing garbage collection or managing compact regions. It also defines various primitive operations, named \emph{external primops}, that expose the RTS capabilities as normal functions. Despite all its responsibilities, however, the RTS is not responsible for the allocation of normal constructors (built in the garbage-collected heap). One reason is that it doesn't have all the information needed to build a constructor heap object, namely, the info table associated to the constructor.

The info table is what defines both the layout and behavior of a heap object. All heap objects representing a same constructor (let's say \mintinline{haskellc}`¤Just`) have the same info table, even when the associated types are different (e.g. \mintinline{haskellc}`Maybe Int` and \mintinline{haskellc}`Maybe Bool`). Heap objects representing this constructor point to a label \mintinline{text}`<ctor>_con_info` that will be later resolved by the linker into an actual pointer to the shared info table.

The RTS is in fact a static piece of code that is compiled once when GHC is built. So the RTS has no direct way to access the information emitted during the compilation of a program. In other words, when the RTS runs, it has no way to inspect the program that it runs and info table labels have long been replaced by actual pointers so it cannot find them itself. But it is the one which knows how to allocate space inside a compact region.

As a result, I need to add two new primitives to GHC to allocate a hollow constructor:

\begin{itemize}
\item one \emph{external primop} to allocate space inside a compact region for a hollow constructor. This primop has to be implemented inside the RTS for the aforementioned reasons;
\item one \emph{internal primop} (internal primops are macros which generates C-{}- code) that will be resolved into a normal albeit static value representing the info table pointer of a given constructor. This value will be passed as an argument to the external primop.
\end{itemize}

All the alterations to GHC that will be showed here are available in full form in~\cite{custom_ghc}.

\paragraph{External primop: allocate a hollow constructor in a region}

The implementation of the external primop is presented in Table~\ref{table:impl-compactAddHollow}. The \mintinline{c}`stg_compactAddHollowzh` function (whose equivalent on the Haskell side is \mintinline{haskellc}`compactAddHollow#`) is mostly a glorified call to the \mintinline{haskellc}`ALLOCATE` macro defined in the \mintinline{text}`Compact.cmm` file, which tries to do a pointer-bumping allocation in the current block of the compact region if there is enough space, and otherwise add a new block to the region.

As announced, this primop takes the info table pointer of the constructor to allocate as its second parameter (\mintinline{c}`W_ info`) because it cannot access that information itself. The info table pointer is then written to the first word of the heap object in the call to \mintinline{c}`SET_HDR`.

\paragraph{Internal primop: reify an info table label into a runtime value}

The only way, in Haskell, to pass a constructor to a primop so that the primop can inspect it, is to lift the constructor into a type-level literal. It's common practice to use a \mintinline{haskellc}`Proxy a` (the unit type with a phantom type parameter) to pass the type \mintinline{haskellc}`a` as an input to a function. Unfortunately, due to a quirk of the compiler, primops don't have access to the type of their arguments. They can, however, access their return type. So I'm using a phantom type \mintinline{haskellc}`InfoPtrPlaceholder# a` as the return type, to pass the contructor as an input!

The gist of this implementation is presented in Table~\ref{table:impl-reifyInfoPtr}. The primop \mintinline{haskellc}`reifyInfoPtr#` pattern-matches on the type \mintinline{haskellc}`resTy` of its return value. In the case it reads a string literal, it resolves the primop call into the label \mintinline{text}`stg_<name>` (this is used in particular to retrieve \mintinline{haskellc}`stg_IND` to allocate indirection heap objects). In the case it reads a lifted data constructor, it resolves the primop call into the label which corresponds to the info table pointer of that constructor. The returned \mintinline{haskellc}`InfoPtrPlaceholder# a` can later be converted back to an \mintinline{haskellc}`Addr#` using the \mintinline{haskellc}`unsafeCoerceAddr` function.

As an example, here is how to allocate a hollow \mintinline{haskellc}`¤Just` constructor in a compact region:

\begin{unbreakable}
{\small
\begin{minted}[linenos]{haskellc}
hollowJust ⩴ Maybe a = compactAddHollow#
  compactRegion#
  (unsafeCoerceAddr (reifyInfoPtr# (# #) ⩴ InfoPtrPlaceholder# 'Just ))
\end{minted}
\vspace{-0.8\baselineskip}
}
\end{unbreakable}

\paragraph{Built-in type family to go from a lifted constructor to the associated symbol}

The internal primop \mintinline{haskellc}`reifyInfoPtr#` that we introduced above takes as input a constructor lifted into a type-level literal, so this is also what \mintinline{haskellc}`fill` will use to know which constructor it should operate with. But \mintinline{haskellc}`DestsOf` have to find the metadata of a constructor in the \mintinline{haskellc}`Generic` representation of a type, in which only the constructor name appears.

So we added a new type family \mintinline{haskellc}`LCtorToSymbol` inside GHC that inspects its (type-level) parameter representing a constructor, fetches its associated \mintinline{haskellc}`DataCon` structure, and returns a type-level string (kind \mintinline{haskellc}`Symbol`) carrying the constructor name, as presented in Table~\ref{table:impl-LCtorToSymbol}.

\begin{table}[p]
\small
\begin{minted}[linenos]{c}
// compactAddHollow#
//   ⩴ Compact# → Addr# → State# RealWorld → (# State# RealWorld, a #)
stg_compactAddHollowzh(P_ compact, W_ info) {
    W_ pp, ptrs, nptrs, size, tag, hp;
    P_ to, p; p = NULL;  // p isn't actually used by ALLOCATE macro
    again: MAYBE_GC(again); STK_CHK_GEN();

    pp = compact + SIZEOF_StgHeader + OFFSET_StgCompactNFData_result;
    ptrs  = TO_W_(%INFO_PTRS(%STD_INFO(info)));
    nptrs  = TO_W_(%INFO_NPTRS(%STD_INFO(info)));
    size = BYTES_TO_WDS(SIZEOF_StgHeader) + ptrs + nptrs;
    
    ALLOCATE(compact, size, p, to, tag);
    P_[pp] = to;
    SET_HDR(to, info, CCS_SYSTEM);
  #if defined(DEBUG)
    ccall verifyCompact(compact);
  #endif
    return (P_[pp]);
}
\end{minted}
\caption{\texttt{compactAddHollow\#} implementation in \texttt{rts/Compact.cmm}}
\label{table:impl-compactAddHollow}
\end{table}

\begin{table}[p]
\small
\begin{minted}[linenos]{haskellc}
case primop of
  [...]
  ¤ReifyStgInfoPtrOp → \_ →  -- we don't care about the function argument (# #)
    opIntoRegsTy $ \[res] resTy → emitAssign (¤CmmLocal res) $ case resTy of
      -- when 'a' is a Symbol, and extracts the symbol value in 'sym'
      ¤TyConApp _addrLikeTyCon [_typeParamKind, ¤LitTy (¤StrTyLit sym)] →
          ¤CmmLit (¤CmmLabel (
            mkCmmInfoLabel rtsUnitId (fsLit "stg_" `appendFS` sym)))
      -- when 'a' is a lifted data constructor, extracts it as a DataCon
      ¤TyConApp _addrLikeTyCon [_typeParamKind, ¤TyConApp tyCon _]
        | ¤Just dataCon ← isPromotedDataCon_maybe tyCon →
          ¤CmmLit (¤CmmLabel (
            mkConInfoTableLabel (dataConName dataCon) ¤DefinitionSite))
      _ → [...] -- error when no pattern matches
\end{minted}
\caption{\texttt{reifyInfoPtr\#} implementation in \texttt{compiler/GHC/StgToCmm/Prim.hs}}
\label{table:impl-reifyInfoPtr}
\end{table}

\begin{table}[p]
\small
\begin{minted}[linenos]{haskellc}
matchFamLCtorToSymbol ⩴ [Type] → Maybe (CoAxiomRule, [Type], Type)
matchFamLCtorToSymbol [kind, ty]
  | ¤TyConApp tyCon _ ← ty, ¤Just dataCon ← isPromotedDataCon_maybe tyCon =
      let symbolLit = (mkStrLitTy . occNameFS . occName . getName $ dataCon)
       in ¤Just (axLCtorToSymbolDef, [kind, ty], symbolLit)
matchFamLCtorToSymbol tys = ¤Nothing

axLCtorToSymbolDef =
  mkBinAxiom "LCtorToSymbolDef" typeLCtorToSymbolTyCon ¤Just
    (\case { ¤TyConApp tyCon _ → isPromotedDataCon_maybe tyCon ; _ → ¤Nothing })
    (\_ dataCon → ¤Just (mkStrLitTy . occNameFS . occName . getName $ dataCon))
\end{minted}
\caption{\mintinline{haskellc}`LCtorToSymbol` implementation in \texttt{compiler/GHC/Builtin/Types/Literal.hs}}
\label{table:impl-LCtorToSymbol}
\end{table}

\section{Evaluating the performance of DPS programming}\label{sec:benchmark}

\paragraph{Benchmarking methodology}

All over this article, I talked about programs in both naive style and DPS style. With DPS programs, the result is stored in a compact region, which also forces strictness i.e. the structure is automatically in fully evaluated form.

For naive versions, we have a choice to make on how to fully evaluate the result: either force each chunk of the result inside the GC heap (using \mintinline{haskellc}`Control.DeepSeq.force`), or copy the result in a compact region that is strict by default (using \mintinline{haskellc}`Data.Compact.compact`).

In programs where there is no particular long-lived piece of data, having the result of the function copied into a compact region isn't particularly desirable since it will generally inflate memory allocations. So we use \mintinline{haskellc}`force` to benchmark the naive version of those programs (the associated benchmark names are denoted with a ``*'' suffix).

\begin{figure}[t]\centering
  \hspace*{-1.5cm}\includegraphics[width=16.8cm]{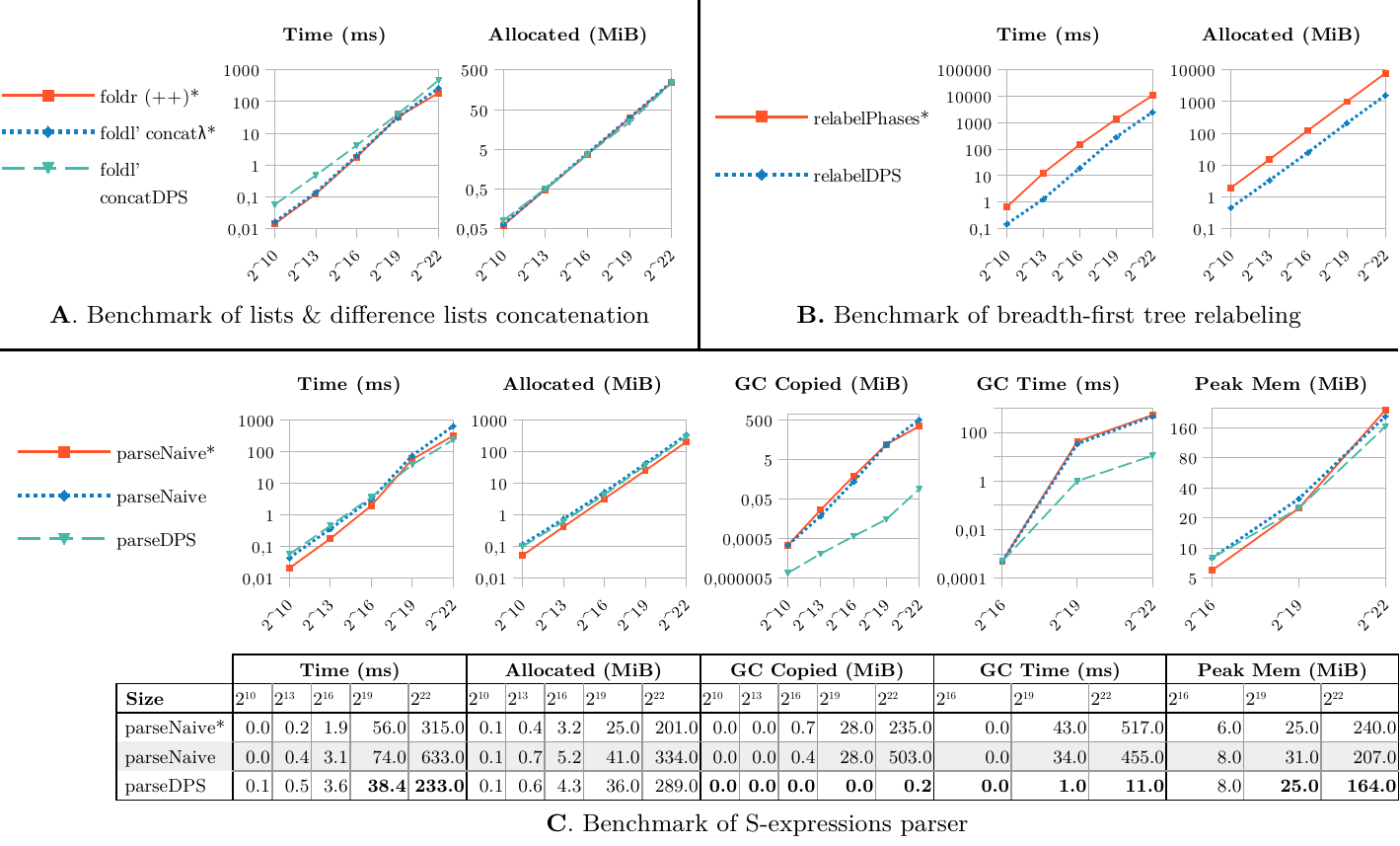}
  \caption{Benchmarks performed on AMD EPYC 7401P @ 2.0 GHz (single core, \texttt{-N1 -O2})}
  \label{fig:bench-charts}
\end{figure}

\paragraph{Concatenating lists and difference lists}

We compared three implementations. \\\mintinline{haskellc}`foldr (++)`* has calls to \mintinline{haskellc}`(++)` nested to the right, giving the most optimal context for list concatenation (it should run in $\mathcal{O}(n)$ time). \mintinline[escapeinside=°°]{haskellc}`foldl' concat°$\lambda$°`* uses function-backed difference lists, and \mintinline{haskellc}`foldl' concatDPS` uses destin\-ation-backed ones, so both should run in $\mathcal{O}(n)$ even if calls to concat are nested to the left.

We see in part \textbf{A} of Figure~\ref{fig:bench-charts} that the destination-backed difference lists have a comparable memory use as the two other linear implementations, while being quite slower (by a factor 2-4) on all datasets. We would expect better results though for a DPS implementation outside of compact regions because those cause extra copying.

\paragraph{Breadth-first relabeling}\label{par:benchmark-bf-tree-traversal}

We see in part \textbf{B} of Figure~\ref{fig:bench-charts} that the destination-based tree traversal is almost one order of magnitude more efficient, both time-wise and memory-wise, compared to the implementation based on \emph{Phases} applicatives presented in~\cite{gibbons_phases_2023}.

\paragraph{Parsing S-expressions}

In part \textbf{C} of Figure~\ref{fig:bench-charts}, we compare the naive implementation of the S-expression parser and the DPS one (see Section~\ref{ssec:parser-sexpr}). For this particular program, where using compact regions might reduce the future GC load of the application, it is relevant to benchmark the naive version twice: once with \mintinline{haskellc}`force` and once with \mintinline{haskellc}`compact`.

The DPS version starts by being less efficient than the naive versions for small inputs, but gets an edge as soon as garbage collection kicks in (on datasets of size $\leq 2^{16}$, no garbage collection cycle is required as the heap size stays small).

On the largest dataset ($2^{22} \simeq 4$MiB file), the DPS version still makes about 45\% more allocations than the starred naive version, but uses 35\% less memory at its peak, and more importantly, spends 47$\times$ less time in garbage collection. As a result, the DPS version only takes 0.55-0.65$\times$ the time spent by the naive versions, thanks to garbage collection savings. All of this also indicates that most of the data allocated in the GC heap by the DPS version just lasts one generation and thus can be discarded very early by the GC, without needing to be copied into the next generation, unlike most nodes allocated by the naive versions.

Finally, copying the result of the naive version to a compact region (for future GC savings) incurs a significant time and memory penalty, that the DPS version offers to avoid.

\section{Related work}

The idea of functional data structures with write-once holes is not new. Minamide already proposed in~\cite{minamide_functional_1998} a variant of $\lambda$-calculus with support for \emph{hole abstractions}, which can be represented in memory by an incomplete structure with one hole and can be composed efficiently with each other (as with \mintinline{haskellc}`fillComp` in Figure~\ref{fig:schema-dlist-concat}). With such a framework, it is fully possible to implement destination-backed difference lists for example.

However, in Minamide's work, there is no concept of destination: the hole in a structure can only be filled if one has the structure itself at hand. On the other hand, our approach introduces destinations, as a way to interact with a hole remotely, even when one doesn't have a handle to the associated structure. Because destinations are first-class objects, they can be passed around or stored in collections or other structure, while preserving memory safety. This is the major step forward that our paper presents.

More recently, \cite{protzenko_mezzo_2013} introduced the Mezzo programming language, in which mutable data structures can be freezed into immutable ones after having been completed. This principle is used to some extend in their list standard library module, to mimic a form of DPS programming. An earlier appearance of DPS programming as a mean to achieve better performance in a mutable language can also be seen in~\cite{larus_restructuring_1989}.

Finally, both \cite{shaikhha_destination-passing_2017} and \cite{bour_tmc_2021} use DPS programming to make list or array processing algorithms more efficient in a functional, immutable context, by turning non tail-recursive functions into tail-recursive DPS ones. More importantly, they present an automatized way to go from a naive program to its tail-recursive version. However, holes/destinations are only supported at an intermediary language level, while both~\cite{minamide_functional_1998} and our present work support safe DPS programming in user-land. In a broader context, \cite{lorenzen_fp_2023} presents a system in which linearity is used to identify where destructive updates can be made, so as to reuse the same constructor instead of deallocating and reallocating one; but this optimization technique is still mostly invisible for the user, unlike ours which is made explicit.

\section{Conclusion and future work}

Programming with destinations definitely has a place in the realm of functional programming, as the recent adoption of \emph{Tail Modulo Cons}~\cite{bour_tmc_2021} in the OCaml compiler shows. In this paper, we have shown how destination-passing style programming can be used in user-land in Haskell safely, thanks to a linear type discipline. Adopting DPS programming opens the way for more natural and efficient programs in a variety of contexts, where the major points are being able to build structures in a top-down fashion, manipulating and composing incomplete structures, and managing holes in these structures through first-class objects (destinations). Our DPS implementation relies only on a few alterations to the compiler, thanks to \emph{compact regions} that are already available as part of GHC. Simultaneously, it allows to build structures in those regions without copying, which wasn't possible before.

There are two limitations that we would like to lift in the future. First, DPS programming could be useful outside of compact regions: destinations could probably be used to manipulate the garbage-collected heap (with proper read barriers in place), or other forms of secluded memory areas that aren't traveled by the GC (RDMA, network serialized buffers, etc.). Secondly, at the moment, the type of \mintinline{haskellc}`fillLeaf` implies that we can't store destinations (which are always linear) in a difference list implemented as in Section~\ref{ssec:dlist}, whereas we can store them in a regular list or queue (like we do, for instance, in Section~\ref{ssec:bf-tree-traversal}). This unwelcome restriction ensures memory safety but it's quite coarse grain. In the future we'll be trying to have a more fine-grained approach that would still ensure safety.

\clearpage{}
\bibliography{bibliography}{}
\bibliographystyle{alpha-fr}
\end{document}